\documentstyle[12pt]{article}

\begin{document} 
\begin{flushright}
Preprint No. IMSC - 93/26
\end{flushright} 
\begin{center}
{\Large\bf{A  CALCULUS  FOR  $SU(3)$ LEADING TO AN ALGEBRAIC 
FORMULA  FOR  THE CLEBSCH-GORDAN COEFFICIENTS}}\\
J.S. Prakash and H.S. Sharatchandra\\
The Institute of Mathematical Sciences,\\
Madras-600 113, India.\\
e-mail: jsp@iopb.ernet.in, sharat@imsc.ernet.in
\end{center}

ABSTRACT:  We  develop a simple computational  tool  for  $SU(3)$ 
analogous to Bargmann's calculus for $SU(2)$.  Crucial new inputs 
are,  (i) explicit representation of the Gelfand-Zetlin  basis  in 
terms  of polynomials in four variables and positive or  negative 
integral  powers of a fifth variable (ii) an  auxiliary  Gaussian 
measure  with  respect  to which the  Gelfand-Zetlin  states  are 
orthogonal  but not normalized (iii) simple generating  functions 
for  generating all basis states and also all invariants.  As  an 
illustration  of  our techniques, an algebraic  formula  for  the 
Clebsch-Gordan  coefficients is obtained for the first time.  This 
involves only Gaussian integrations.    Thus   $SU(3)$  is  made   
as   accessible   for computations as $SU(2)$ is.\\
\newpage
\section{Introduction}
Compact  Lie groups have been extensively studied from  different 
viewpoints\cite{BAO,ZDP,WBG1,GR,HM,MFD2,WEP,WH,RG1,BH,LDE,LDB}. 
In spite  of  this,  there  are  gaps  in   our 
understanding  which  are keenly felt in  specific  applications. 
This  has  mostly to do with the absence of a  viable  scheme  of 
general computations.  For example, there is no algebraic formula for the 
Clebsch-Gordan coefficients of even the $SU(3)$ group, in spite of 
extensive  work by a generation of mathematical physicists.  This 
is  in contrast to $SU(2)$ group, where it seems that  everything 
can  be computed in more than one way.  Somehow, every  technique 
that  works  for  $SU(2)$  does  not  appear  to  have  a  simple 
generalization for other groups.

     In  this paper we develop techniques which provide  a  simple 
computation   tool   for $SU(3)$.  Our   aim  is   to   highlight  
the   flexibility   available    for  computations   of   various 
objects  of interest  in  representation theory.In particular  we 
obtain  a  closed 
formula  for the Clebsch-Gordan coefficients of $SU(3)$.  With  a 
couple of new inputs it might be possible to use our techniques 
for  other groups also.  We have borrowed ideas heavily from  many 
earlier  workers.  We  have made some  conceptual  and  technical 
advances which together have enabled us to provide a simple  tool. 
    
 We  now give a summary of the earlier works,  with  specific 
reference to $SU(3)$.  An excellent summary of the situation up to 
1971  is contained in the Appendix of the review by  Smorodinskii 
and  Shelepin $\cite{SY}$.  Our summary is by no  means  complete 
and accurate.  

     In $SU(2)$ \cite{SY,BLC9} case there are broadly three 
computational tools. 
(a) infinitesimal approach, (b) polynomial basis and  generating 
invariants, (c) use of relationship with the symmetric group.
Within each approach there have been many different ways \cite{SY,BLC9} of 
deriving  formulae  for the Clebsch-Gordan  coefficients.  Almost 
every one of these variations has been tried for $SU(3)$, but each 
has led to obstacles.

     A  major obstacle encountered in any approach is  the  outer 
multiplicity  problem.  In the decomposition into irreducible 
representations 
(IRs) of a Kronecker product  of  two  IRs  of $SU(n)$,  $n>3$,  a  
given IR may appear more  than  once.  These repeating  IRs cannot 
be distinguished by the matrix elements  of the generators.  We need 
to understand how the repeating IRs  may be  distinguished,  labeled  
in a  convenient  and canonical way  and  handled.  
Extensive  efforts have been put into this problem.   At least  in 
the  $SU(3)$ case, the problem has been essentially  resolved  by 
many  authors  using  diverse techniques and  often  without  the 
knowledge  of  previous works.  New  Casimirs  (`Chiral  Casimirs') 
which  distinguish  between repeating IRs have  been  constructed 
\cite{MM1,MAJ,GGH}.  Biedenharn and his collaborators have  obtained 
a 'canonical  resolution' of the multiplicity problem \cite{BLC7,BLC8}.  
An explicit formula for multiplicity has been obtained by Coleman 
\cite{PB}  by  analyzing the  Littlewood-Richardson  rule;  Jasselette 
\cite{JP1,JP2}  and  Resnikoff  \cite{RM}  by  applying  the  theory   of 
invariants;  O'Reilly \cite{OMF} by a detailed and careful  analysis  of 
the   Kronecker   product.  Anishetty,    Gadiyar,   Mathur   and 
Sharatchandra  \cite{AGMS}  recently reinterpreted these  results  to 
give  the most explicit formula, analogous to the  triangle  rule 
for  the addition of angular momenta in $SU(2)$.  The  advantage 
of  this  formula is that the IRs in the  Kronecker  product  are 
labeled by ${\mbox{\it free}}$ integers which are subject  only 
to  additive constraints by the two IRs one started  with.   This 
therefore  provides  a 'natural labeling' of  the  repeating  IRs. 
Gadiyar   and   Sharatchandra  \cite{GS}  have  recently   solved   the 
multiplicity  problem for $SU(n)$ for any $n$.  This is done  by 
obtaining  an  explicit  algebraic solution  of  the  Littlewood-
Richardson rule in terms of free integers.

     In the infinitesimal approach to $SU(2)$, the Clebsch-Gordan 
coefficients  are  computed as follows. A recursion  relation  is 
obtained  by considering the action of the Lie algebra on the  direct 
product space.  This does not work as it is for $SU(3)$ and other 
groups.   The  Lie  algebra does not  provide  enough  number  of 
recursion relations to be able to compute all Clebsch-Gordan 
coefficients. The reason is the multiplicity problem. Biedenharn 
and collaborators \cite{BLC5,BLC6} have emphasized the need to define 
a  basic  set  of  irreducible tensor  operators.  The  set  they 
construct  provides a 'canonical resolution' of the  multiplicity 
problem.   For a review see \cite{BLC8,LJD1,LJD2}. Their  'pattern  
calculus' 
provides   a   framework   for   computing   the   Clebsch-Gordan 
coefficients.  There  has  been extensive  formal  work  in  this 
direction.  It has led to significant concepts such as  a  global 
algebraic formulation of $SU(3)$ tensor operator structure \cite{BLC5} 
and   the   denominator  functions  \cite{BLC6}   which   have   wider 
ramifications  \cite{GIM5,BLC4,BAJ1,BAJ2,DJ}.  In  addition,  this 
approach has been very useful for practical algorithms \cite{DJP,LB1,LB2}  
and  symbolic  manipulation programs  \cite{KWH}.  However  the 
approach  has  not  (yet!) led to an algebraic  formula  for  the 
Clebsch-Gordan coefficients of $SU(3)$.

     The   starting   point  of  the  second  approach   is   the 
construction of a convenient {\it{ model  space }} i.e. a concrete 
realization  of  (say, on a function space,) the  basis  of  every 
irreducible  unitary  representations of the group.  In  case  of 
$SU(2)$, the simplest realization of the basis is as polynomials 
in  two  complex variables. This was known to Weyl \cite{WH}  and  was 
used  by van der Waerden \cite{vdW}, Cartan \cite{CE} and Kramers 
\cite{KHA,BHC}.   It is  related to the spinor representation of $SO(3)$.  
Schwinger's 
\cite{SJ}  boson  calculus  is also related  to  this.  This  approach 
reached  the  peak  in  the work of  Bargmann  \cite{BV,KHA}  where  all 
computations  in  $SU(2)$ are reduced to evaluation  of  Gaussian 
integrals. The computation of Clebsch-Gordan coefficients amounts 
to  the construction of invariant polynomials. We will refer  to 
this package of tools as the Bargmann calculus. We give a constructive 
analysis of this calculus in Sec.2.  Though there are other model spaces 
for $SU(2)$, eg. the spherical harmonics, none provide as simple a 
computational tool.

     There  has  been extensive work to  generalize  this  second 
approach   to   other  groups.  Many  model  spaces   have   been 
constructed.  Realization using polynomials 
\cite{MM1,BIN,ZDP,JP1,RM,KVP,PJS2} boson calculus 
\cite{BAJ1,LE,BAJ2} harmonic functions 
i.e.  functions  on  coset spaces \cite{BMA,KD,GIM6}.  Gelfand  and 
collaborators\cite{BIN} have obtained a differential  equation  which 
yields the measure with respect to which the Gelfand-Zetlin basis 
states  are orthonormal. Jasselette \cite{JP1,JP2} Resnikoff \cite{RM}, 
and  Karasev \cite{KVP} have constructed invariant  polynomials  from 
which the Clebsch-Gordan coefficients may be obtained in principle. 
Resnikoff  \cite{RM} made progress in using a Gaussian measure  to  extract 
the Clebsch-Gordan coefficients.

     In spite of all this work, the situation is not comparable to 
the  $SU(2)$ case. Some of the coefficients \cite{RM,JP2,KVP} in  the 
formula for the Clebsch-Gordan coefficients cannot be  explicitly 
computed.  The stumbling blocks in this approach which make it so 
much harder to handle $SU(3)$ are the following. The realization 
of the basis functions in terms of polynomials is much more 
complicated. In fact all earlier realizations
\cite{MM1,PJS2} are analogous to the harmonic polynomials (i.e. 
those which are
annihilated by the Laplacian operators) obtained from the defining
representation of $SO(3)$ rather than to the monomials obtained 
from the spinor representation. Explicit construction of such basis 
vectors \cite{MM1,LE} and
working with them is not easy. Moreover the measure with respect 
to which the basis is orthonormal is not known in a closed form 
\cite{BIN}. Even were it 
known explicitly, the hope of computing with it appears remote. 
Invariant polynomials in the space of three IR's can be easily 
built \cite{JP1,RM,KVP}.  However the invariant polynomial 
consistent with the three given IR's is not unique in general. 
The coefficients have to be fixed by demanding that they 
be expandable in terms of the constrained polynomials representing 
the basis vectors. Even this does not fix the invariant polynomial 
completely. This is a consequence of the multiplicity problem and 
requires a choice of basis to be made in the space of repeating 
IR's. After all this, there is no easy way of
extracting the Clebsch-Gordan coefficients \cite{MM1,JP1,JP2,RM,KVP}.

	In this paper we show how these stumbling blocks may all be 
overcome. We
develop a calculus which is almost as simple as Bargmann's calculus. All
computations are effectively reduced to Gaussian integrations.

	The first simplification we have achieved is in the explicit 
realization
\cite{PJS1} of the Gelfand-Zetlin basis vectors, free of constraints. 
We realized this in Sec.3. Our realization uses polynomials in four 
complex variables and positive or negative integral powers of a fifth 
variable. It is related to the functions on the cone 
$\vec{w}\cdot \vec{z} = 0$ where $\vec{z}$ and $\vec w$ each are triplets 
of complex variables \cite{ZDP,PJS2}.  We choose a specific parameterization  
of the cone. (i.e. eliminating $w_3$) and explicitly construct 
Gelfand-Zetlin basis for the functions on the space. With our 
parameterization, we are using 
all polynomials and not just a subset as in earlier works.

 Our realization is not as simple as the monomial basis for
$SU(2)$. However, in Sec.4 we use a generating function which 
generates all the (unnormalized) basis functions of every IR. This 
generating function is as
simple as the "principal vectors" of Bargmann's calculus.

	At this stage of our formulation the normalizations of our basis 
vectors are not known. The normalization is to be determined by requiring 
that the representation matrix for each IR be unitary. It is always a 
headache to compute the normalization \cite{MM1,LE}.  The great advantage 
of Bargmann's calculus is the Gaussian measure which permits explicit and 
easy computations. It is fortuitous that in the $SU(2)$ case the measure 
with respect to which the Gelfand-Zetlin  states are orthonormal  is so 
simple.  In order to retain this computability, we construct an auxiliary 
Gaussian measure with respect to which the Gelfand-Zetlin basis vectors 
are orthogonal but are not automatically normalized (Sec.5). In fact we use 
this measure to compute the normalization itself by requiring that the 
representation matrix in each IR be unitary (Sec.6).
	This way we are killing two birds with one stone: To start 
with, the normalization  and the measure are both unknown. A simple 
auxiliary measure is
constructed and used to compute the normalization itself.
The basis we use also leads to a simplification in the form of 
invariants in the direct product space of three IR's (Sec.8). This 
is a consequence of using 
the cone $\vec w\cdot \vec z =0$. The invariant polynomial corresponding to 
a choice of a repeating IR is now uniquely known and there are no unknown 
coefficients to be fixed separately (Sec.8). Therefore the Clebsch-Gordan 
coefficients (Sec.7) can be obtained by simply expanding this 
polynomial in the basis vectors of each IR. We are assured that 
such an expansion exists 
because our basis spans all polynomials in contrast to the constrained 
polynomials of the earlier works.

	In $SU(2)$ case, the Clebsch-Gordan coefficients are simply 
obtained by reading off the coefficients of the right monomial in the 
invariant polynomial. But our basis is more complicated. In order to 
obtain the coefficients in the expansion, we again use our auxiliary 
measure.  We introduce a generating function for the invariant polynomials 
themselves (Sec.9). This way all Clebsch-Gordan coefficients are being 
computed in one shot. Moreover, it is easier to do these computations 
than with each invariant polynomial individually. The Gaussian measure 
is used to find the inner product of the generating function for the 
basis vectors of the three IR's with the generating function for the 
invariant polynomials (Sec.10).  There are terms in the exponent which 
are ${\mbox{\it apparently}}$ cubic in this integration -- a consequence 
of the "multiplicity problem". Remarkably however, because of the specific 
measure we have chosen, the integrals can all be explicitly computed.

\section{An analysis of Bargmann's technique for $SU(2)$}
Bargmann \cite{BV} used an axiomatic approach in his analysis.  This 
presumes many results known from other methods.  In this section we 
give a constructive analysis of  Bargmann's techniques for $SU(2)$.  
This will set the stage for our techniques for $SU(3)$, making it 
clear as to where new ideas are required.
$SU(2)$ is the group of simple unitary $2 \times  2$ matrices.  Its action
on ${\cal C}_2$, the $2$-dimensional complex Euclidean space, is given by
\begin{eqnarray}
\pmatrix{z_1 \cr z_2}\rightarrow {\cal U}\pmatrix{z_1 \cr z_2} 
\label{eq:3.2.1}
\end{eqnarray}
where ${\cal U} \in SU(2)$.
The doublet of complex numbers $(z_1, z_2)$ transforms as the irreducible
representation (IR) $\underline2$ of $SU(2)$.  In particular, $z_1$ 
represents the spin 'up' state and $z_2$, the spin 'down' state.  
States of an arbitrarily high spin can be obtained from a large enough 
collection of spin $1/2$ particles.  In particular, states of spin
$J$ can be obtained from a system of $2J$ number of identical spin 
$\frac{\displaystyle 1}{\displaystyle 2}$
particles,  i.e. from $2J$ copies of ${\cal C}_2$. ($2J$ boxes  in 
the row of Young tableaux represent spin $J$). This corresponds to 
a realization of the IRs in the space of polynomials in $z_1$ and 
$z_2$ :  the monomial
\begin{eqnarray}
z^m_1 z^n_2
\label{eq:3.2.2}
\end{eqnarray}
describes, up to a normalization, the basis states
\begin{eqnarray}
\vert{JM}>, \quad 2J = m + n, \qquad  2M = m - n   
\label{eq:3.2.3}
\end{eqnarray}

  We notice that as $m$ and $n$ range independently over all non-negative
integers, every basis state $\vert{JM}>$ of every IR is realized
uniquely.  This means the following.

Consider the space ${\cal F}_2$ whose elements are, roughly speaking,
polynomials in $z_1$ and $z_2$.  In this space, every IR is realized, and
moreover each IR is realized once only.  Thus it is a model space (see 
Sec.1 for the definition).  In
addition, the standard basis states(eq:~\ref{eq:3.2.3}) are simply 
realized as monomials.  Thus this space is very convenient for 
calculations.  There are further surprises to follow.  The action of 
the group on any state is obtained in this
model by transforming $z_1$ and $z_2$ in (eq:~\ref{eq:3.2.2}) as given by 
(eq:~\ref{eq:3.2.1}).  To express
this action on an arbitrary state, it is very convenient to work with 
the generating function,
\begin{eqnarray}
{\cal Z}(a, b) = exp(az_1 + bz_2)      
\label{eq:3.2.4}
\end{eqnarray}
Simply by extracting the coefficient of $a^m b^n$, an unnormalized 
basis state (eq:~\ref{eq:3.2.3}) can be extracted.  This way, we are 
handling all states of all IRs in one shot.  Moreover, the action of 
the group on the generating function is very simple:
\begin{eqnarray}
{\cal U}: {\cal Z}(a, b) \rightarrow {\cal Z}\left ((a, b){\cal U}\right )
      \label{eq:3.2.5}
\end{eqnarray}
 The normalizations of the basis states (within an IR) are obtained by 
demanding the unitarity of the representation matrices on the space.  It is
sufficient to use group elements close to the identity for this purpose. An 
$SU(2)$ matrix close to identity may be represented as follows:
\begin{eqnarray}
{\cal U} \quad \simeq \quad 1 + i(\epsilon_3\sigma_3 + \epsilon_+ \sigma_+ 
+ \epsilon_- \sigma_-)
\label{eq:3.2.6}
\end{eqnarray}
where, 
\begin{eqnarray}
\sigma_3 = \pmatrix{1 & 0 \cr 0 & -1};                    
\quad \sigma_+ = \pmatrix{0 & 1 \cr 0 & 0};  
\quad \sigma_- = \pmatrix{0 & 0 \cr 1 & 0}
\label{eq:3.2.7}      
\end{eqnarray}
(Note that we are not using conventional normalization for the generators).

For unitarity, ${\cal U}^\dagger {\cal U} = 1$, we require
\begin{eqnarray}
\epsilon^*_+ = \epsilon_-, \quad \epsilon^*_3 = \epsilon_3
\label{eq:3.2.8}
\end{eqnarray}
(det ${\cal U} = 1$ is satisfied because the matrices (eq:~\ref{eq:3.2.7}) 
are traceless).  Using
(eq:~\ref{eq:3.2.6}) in (eq:~\ref{eq:3.2.5}), we get, the following 
representation for the generators on
our model space,
\begin{eqnarray}
\pi^0 = a\frac{\partial}{\partial a} - b\frac{\partial}{\partial b}; 
\quad \pi^- = a\frac{\partial}{\partial b}; 
\quad \pi^+ = b\frac{\partial}{\partial a}   
\label{eq:3.2.9}
\end{eqnarray}
where the generators correspond to $\sigma_3$, $\sigma_+$ and $\sigma_-$
respectively.  The notation we are using is motivated by the isospin 
triplet of pions.
The requirement of unitarity of the representation matrix translates 
into the following condition on the generators:
\begin{eqnarray}
(\pi^0)^* = \pi^0, \quad (\pi^+)^* = \pi^-
\label{eq:3.2.10}
\end{eqnarray}
where $*$ stands for the adjoint.
We write formally,
\begin{eqnarray}
{\cal Z}(a, b) = \sum_{m, n} {a^m b^n \vert{m, n}})
\label{eq:3.2.11}
\end{eqnarray}
where, 
\begin{eqnarray}
\vert{m, n}) = \frac{z^m_1 z^n_2}{m!n!}     
\label{eq:3.2.12}
\end{eqnarray}
represent unnormalized basis states.
The normalized basis states 
\begin{eqnarray}
\vert{J, M}> \equiv \vert{m, n}> = N^{-1/2}(m, n)\vert{mn}), 
\quad 2J = m + n \quad 2M = m - n  
\label{eq:3.2.13}
\end{eqnarray}
are to be obtained by requiring,
\begin{eqnarray}
<{m'n'}\vert{T^*}\vert{mn}> = <{mn}\vert{T}\vert{m'n'}>^* 
\label{eq:3.2.14}
\end{eqnarray}
for every generator $T$.  Let
\begin{eqnarray}
T\vert{mn}) = \sum_{m'n'}T(mn;m'n') \vert m'n')
\label{eq:3.2.15}
\end{eqnarray}
This action can be easily computed using the generating function 
(eq:~\ref{eq:3.2.4}) and the
representation (eq:~\ref{eq:3.2.9}) of the generators.  In terms of 
normalized states this means, 
\begin{eqnarray}
N^{1/2}(m,n)T\vert{mn}> =
\sum_{m'n'}T(mn;m'n')N^{\frac{1}{2}}(m',n')\vert m'n'> 
\label{eq:3.2.16}
\end{eqnarray}
Using orthonormality,
\begin{eqnarray}
<{m'n'}\vert{mn}> = \delta_{m'm}\delta_{n'n}
\label{eq:3.2.17}
\end{eqnarray}
We get
\begin{eqnarray}
<{m'n'}\vert{T}\vert{mn}> =
T(mn;m'n')\frac{N^\frac{1}{2}(m',n')}{N^\frac{1}{2}(m,n)} 
\label{eq:3.2.18}
\end{eqnarray}
Define $T^*(mn;m'n')$ for the generator $T^*$, analogously to 
(eq:~\ref{eq:3.2.15}).  Condition
(eq:~\ref{eq:3.2.14}) gives, 
\begin{eqnarray}
\left | {\frac{N(m,n)}{N(m'n')}}\right | = \frac{T(mn;m'n')}{T^*(m'n';mn)}
\label{eq:3.2.19}
\end{eqnarray}
This way, the relative normalizations of basis states within an IR may 
be computed.  For the present case,
\begin{eqnarray*}
\pi^0 \sum{a^mb^n\vert{mn}})=\sum{(m-n)a^mb^n\vert{mn}})
\end{eqnarray*}
\begin{eqnarray*}
\pi^- \sum{a^mb^n\vert{mn}})=\sum{a^{m+1}b^{n-1}n\vert{mn}})
\end{eqnarray*}
\begin{eqnarray}
\pi^+ \sum{a^mb^n\vert{mn}})=\sum{a^{m-1}b^{n+1}m\vert{mn}})
\label{eq:3.2.20}
\end{eqnarray}
Comparing like powers of $a$ and $b$, we get,
\begin{eqnarray*}
{\pi^0(mn;mn) = m-n}      
\label{eq:3.2.21a}
\end{eqnarray*}
\begin{eqnarray*}
{\pi^-(mn;m-1, n+1) = (n+1)} 
\label{eq:3.2.21b}
\end{eqnarray*}
\begin{eqnarray}
{\pi^+(mn;m+1, n-1) = (m+1)}  
\label{eq:3.2.21c}
\end{eqnarray}
Other matrix elements are zero.  
$\pi^0$ in (eq:~\ref{eq:3.2.19}) does not lead to any constraints on the 
normalizations.  This is because it is diagonal in the chosen basis.  
However, using (eq:~\ref{eq:3.2.19}) for
$\pi^{\pm}$, we get, 
\begin{eqnarray}
\left | {\frac{N(m,n)}{N(m-1,n+1)}}\right |  = \frac{n+1}{m} 
\label{eq:3.2.22}
\end{eqnarray}
We choose the solution,
\begin{eqnarray}
N(m, n) = \frac{1}{m!n!}  
\label{eq:3.2.23}
\end{eqnarray}
The solution is determined only up to (i) any function of the sum 
$m+n = 2J$, (ii) an arbitrary phase factor. (i) means that the 
relative normalization of states in different IRs is not fixed
by our criterion.  This is to be expected because unitarity of the
representation matrix constrains relative normalizations of the basis 
states only within each IR.  For any phase, unitarity is assured, and 
corresponds to a choice of the phases of the basis states. 

Our orthonormalized basis states are represented by 
\begin{eqnarray}
\vert{mn}> \quad = \quad  \frac{z^m_1z^n_2}{\sqrt{m!}{\sqrt{n!}}}
\label{eq:3.2.24}
\end{eqnarray}
\indent It would be easy to guess the measure with respect to which this basis is 
orthonormal.  Define the inner product:
\begin{eqnarray}
(f, g) =
\int{\frac{d^2z_1}{\pi}\frac{d^2z_2}{\pi}}exp(-\bar{z}_1z_1 -
\bar{z}_2z_2)\overline{f(z_1,z_2)}g(z_1,z_2) 
\label{eq:3.2.25}
\end{eqnarray}
for functions in ${\cal F}_2$.  The states (eq:~\ref{eq:3.2.24}) are 
orthonormal with respect to
this measure.  Note that the measure is invariant under the action of the 
group (eq:~\ref{eq:3.2.1}).  We are led to a simple, Gaussian measure. As a 
consequence, it is easy to obtain a general formula for the Clebsch-Gordan 
coefficients. We now review this method.

Consider the direct product of two IRs, $J_1$ and $J_2$.  This 
representation of the group is reducible in general.  Consider its
decomposition into various irreducible components:
\begin{eqnarray}
\vert{(J_1J_2)J_3M_3>}  = \sum_{M_1, M_2}C^{J_1 \, \,J_2 \,\,\,\, J_3}_
{M_1 M_2 M_3} \vert J_1 M_1>\vert J_2 M_2>   
\label{eq:3.2.26}
\end{eqnarray}
The coefficients in the expansion are the Clebsch-Gordan coefficients.  It 
is more convenient
and symmetric to write this as follows.  Given two IRs, $J_1$ and $J_2$, 
it is possible to form a non-trivial combination invariant under the group, 
only if $J_1 = J_2$ and in this case, 
\begin{eqnarray}
\sum (-1)^{(J-M)} \vert{J, M}> \vert{J, -M}> = \mbox{invariant}
\label{eq:3.2.27}
\end{eqnarray}
As a consequence (eq:~\ref{eq:3.2.26}) may be reinterpreted as follows.  
Given three IRs, $J_1$,
$J_2$ and $J_3$, try and form a (non-trivial) invariant combination. 
\begin{eqnarray}
\sum_{M_1, M_2, M_3}\pmatrix{J_1 & J_2 & J_3 \cr M_1 & M_2 &
M_3}\vert{J_1M_1}>\vert{J_2M_2}>\vert{J_3M_3}>
\label{eq:3.2.28}
\end{eqnarray}
The coefficients are the $3-j$ symbols.

Now represent the three IRs by homogeneous polynomials of degrees 
$2J_1$, $2J_2$,
$2J_3$ in variables ${(z^1_1, z^1_2)}$, ${(z^2_1, z^2_2)}$ and 
${(z^3_1, z^3_2)}$, respectively.
Then (eq:~\ref{eq:3.2.28}) corresponds to forming an invariant combination 
out of such polynomials.  It is easy to do this.
Invariant theory implies that any invariant polynomial in the six 
variables is a polynomial in the three independent invariants,
\begin{eqnarray}
(z^1_1z^2_2-z^1_2z^2_1),~~~~ (z^2_1z^3_2-z^2_2z^3_1),~~~~ 
(z^3_1z^1_2-z^3_2z^1_1) 
\label{eq:3.2.29} 
\end{eqnarray}
In order to satisfy our homogeneity requirements, we need non-negative
integers $N_1, N_2$ and $N_3$ in :
\begin{eqnarray}
{(z^1_1z^2_2-z^1_2z^2_1)}^{N_3} {(z^2_1z^3_2-z^2_2z^3_1)}^{N_1} 
{(z^3_1z^1_2-z^3_2z^1_1)}^{N_2} 
\label{eq:3.2.30}
\end{eqnarray}
such that 
\begin{eqnarray}
2J_1 = N_2 + N_3, \quad 2J_2 = N_3 + N_1, \quad 2J_3 = N_1 + N_2
\label{eq:3.2.31}
\end{eqnarray}
For given $J_1, J_2$ and $J_3$, the only solution is, 
\begin{eqnarray*}
N_1=J_2+J_3-J_1
\end{eqnarray*}
\begin{eqnarray*}
N_2=J_3+J_1-J_2
\end{eqnarray*}
\begin{eqnarray}
N_3=J_1+J_2-J_3
\label{eq:3.2.32}
\end{eqnarray}
if the right hand side are all non-negative.  Thus if $J_1$,$J_2$ and 
$J_3$ satisfy
the triangle condition, there is a unique invariant.  Otherwise there 
is no non-trivial invariant.
Thus the $3-j$ symbols are obtained (up-to an overall normalization 
depending only on the total spins $J_1$, $J_2$ and $J_3$) by simply 
extracting in (eq:~\ref{eq:3.2.30}) the
coefficients of the monomials (eq:~\ref{eq:3.2.24}) in the three sets of 
variables.

To obtain a formula for the $3 - j$ symbols, we have to apply the 
binomial  theorem,  and  extract  the  relevant  powers  of   the 
monomials.  We get, (up-to a normalization), 
\begin{eqnarray}
\sum_{p_a,q_a}\prod^3_{a=1}\frac{(-1)^{q_a}}{p_a!q_a!}
\label{eq:32A}
\end{eqnarray}
where  the sum is over all non  -  negative 
integers ${p_a, q_a}$ satisfying the following matrix equation.
\begin{eqnarray}
\pmatrix  {N_1  &  N_2 & N_3\cr m^1 & m^2 & m^3\cr n^1  &  n^2  & 
n^3}\quad  = \pmatrix{q_1 + p_1 & q_2 + p_2 & q_3 +  p_3\cr q_2 + 
p_3 & q_3 + p_1 & q_1 + p_1 \cr
q_3 + p_2 & q_1 + p_3 & q_2 + p_1}
\label{eq:32B}
\end{eqnarray}
\section{A Model Space for $SU(3)$}
	Our first task is to construct a convenient model space for $SU(3)$ 
(model space has been defined in Sec. 1). It is
not possible to get a model space as simple as the one for $SU(2)$. But, we
have constructed \cite{PJS2} a model space which is simple enough for 
obtaining general formulae.  We provide an ab initio review of this 
construction in this section.

	In case of $SU(2)$ all IRs could be constructed from the defining
representation $\underline{2}$. This is no longer true for other semi-simple
groups.Consider a triplet $(z_1, z_1, z_3)$ of complex numbers transforming 
as the defining representation $\underline{3}$ of $SU(3)$.  By considering 
polynomials in these complex variables we can only build totally  
symmetric tensors of $SU(3)$.
Such IRs are represented by Young's tableaux with just our row. A general
Young's tableau has two rows, some columns having  two boxes and the rest
having  one box.  In order to build a general IR, observe that an IR 
with one column of two rows corresponds to the $\underline{3^*}$ of 
$SU(3)$.  Therefore a general IR can be built using a direct product of 
$\underline{3^*}$s and $\underline{3}$s.  Further, the tensors 
corresponding to the Young's tableaux
are symmetric in indices along each row.  This means that it suffices to
consider direct products which are totally symmetric in the 
$\underline{3^*}$s and in the $\underline{3}$s.  Therefore, we may 
build a general IR in the space of polynomials in two triplets of 
complex numbers $(z_1, z_2, z_3)$ and $(w_1, w_2, w_3)$ transforming 
as    $\underline{3}$   and   $\underline{3}^*$    of    $SU(3)$, 
respectively.  (All   this  is  a  heuristic  explanation  of   a  
result  proven  in \cite{BAO,ZDP}).
 
	IRs of $SU(3)$ are conveniently labeled by two arbitrary 
non-negative integers $(M, N)$ which stand for the number of columns 
with one box and two boxes, respectively.  Such an IR can be realized 
using polynomials of degree $M$ in the $z$'s and $N$ in the $w$'s, 
i.e.,  polynomials built from the monomials,

\begin{eqnarray}
z^{m_1}_1 z^{m_2}_2 z^{m_3}_3 w^{n_1}_1 w^{n_2}_2 w^{n_3}_3 
\label{eq:3.1} 
\end{eqnarray}
with,
\begin{eqnarray}
m_1 + m_2 + m_3 = M, \qquad n_1 + n_2 + n_3 = N
\label{eq:3.2}
\end{eqnarray}
However, this space contains some other IRs $(M', N')$ with 
$M' < M$ and $N' < N$.  The reason is that it is possible to form an 
$SU(3)$ invariant $\vec w\cdot\vec  z$.  This is again a major 
difference from the $SU(2)$ case.  A simple way to remove
the unwanted IRs is to impose the constraint,
\begin{eqnarray}	
\vec w\cdot \vec z = 0
\label{eq:3.3}
\end{eqnarray}			
where,
\begin{eqnarray}	
\vec w\cdot \vec z = w_1 z_1 + w_2 z_2 + w_3 z_3
\label{eq:3.4}
\end{eqnarray}			
(i.e we are constraining our variables to a cone in $C_6$).  The IR
$(M, N)$ is now uniquely realized in the subspace with constraints 
(eq:~\ref{eq:3.2}).

	We are forced to contend with the constraint (eq:~\ref{eq:3.3}) 
in order to get a model space.  We obtain an explicit and simple 
enough basis by simply eliminating
$w_3$ (say), in favor of the other five variables.
\begin{eqnarray}
w_3 = - \frac{1}{z_3} ( w_1 z_1 + w_2 z_2)
\label{eq:3.5}
\end{eqnarray}			
	Thus our space is spanned by the monomials, (allowing for 
negative powers of $z_3$)
\begin{eqnarray}
z^{m_1}_1 z^{m_2}_2 w^{n_1}_1 w^{n_2}_2 (w_1z_1+w_2z_2)^{n_3} z^{m_3-n_3}_3
\label{eq:3.6}
\end{eqnarray}

	In order to get an explicit realization of the Gelfand - Zetlin
basis in this space, we proceed as follows.  Note that $(z_1, z_2)$ 
transforms as a $\underline2$ and $(w_1, w_2)$ as a $\underline2^*$ 
(which is equivalent to $\underline2$) under the isospin $SU(2)$ 
subgroup of $SU(3)$.  The combination $(w_1z_1 + w_2z_2)$ in (3.6) is 
an $SU(2)$ singlet built of these two doublets.  This suggests that it 
is useful to use the coupled basis for the isospin group.
This is done as follows.  The monomials $z^{m_1}_1z^{m_2}_2$ and
$w^{n_1}_1w^{n_2}_2$ with
\begin{eqnarray}
m_1 + m_2 = 2I' \qquad  n_1 + n_2 = 2I''  
\label{eq:3.7}
\end{eqnarray}
span the IRs of (iso)spin $I'$ and $I''$, respectively.  Therefore the 
direct product of these two spaces is a direct sum of spaces with 
isospin $I' + I''$, $I' + I'' - 1$,$\ldots$ , $\vert I' - I''\vert $, 
each isospin appearing just once.  This decomposition can be performed 
explicitly by the following trick : Introduce an (external) doublet 
$(p,q)$ transforming as a $\underline2^*$ of
$SU(2)$.  Then the following combination is invariant under $SU(2)$:

\begin{eqnarray}
(pz_1 + qz_2)^R (pw_2 - qw_1)^S (w_1z_1 + w_2z_2)^T
\label{eq:3.8}
\end{eqnarray}
Now,
\begin{eqnarray}
\frac{p^P q^Q}{\sqrt{P!}\sqrt{Q!}} \sim \, \vert I = \frac{P+Q}{2}, 
\, I_3 = \frac{-P+Q}{2} >                      
\label{eq:3.9}
\end{eqnarray}
under $SU(2)$ transformations.  On using (eq:~\ref{eq:3.2.1}) this means 
that the coefficient of the monomial $p^Pq^Q$ corresponds to the state,
\begin{eqnarray}
\vert (I'\, I'') I\, I_3 >
\label{eq:3.10}
\end{eqnarray}
of the coupled basis, where
\begin{eqnarray*}
R + S = P + Q = 2I
\end{eqnarray*}
\begin{eqnarray*}
S + T = n_1 + n_2 = 2I''
\end{eqnarray*}
\begin{eqnarray*}
T + R = m_1 + m_2 = 2I'
\end{eqnarray*}
\begin{eqnarray}
P - Q = 2I_3 
\label{eq:3.11}		
\end{eqnarray}	
	This way, we are able to explicitly construct basis vectors of 
the coupled basis.  By allowing for all non-negative integer values of 
$P,Q,R,S$ and $T$ subject to the constraints of (eq:~\ref{eq:3.11}) we 
are simply making a change of basis from the basis (eq:~\ref{eq:3.7}).  
We make this change of basis in the space spanned by (eq:~\ref{eq:3.6}) 
(further constrained by (eq:~\ref{eq:3.7})).  We get an equivalent basis 
(as coefficients of $p^Pq^Q$)in,

\begin{eqnarray}
(pz_1+qz_2)^R (pw_2 - qw_1)^S (w_1z_1 + w_2z_2)^{T+n_3} z^{m_3-n_3}_3
\label{eq:3.12}
\end{eqnarray}

In terms of the new parameters we have,
\begin{eqnarray}
R + T + m_3 = M  \qquad  S + T + n_3 = N  
\label{eq:3.13}
\end{eqnarray}
	We now notice that distinct values of $m_3$, $n_3$ and $T$ such 
that $T + n_3$ and $T + m_3$ have same values correspond to the same 
basis vector on the cone $\vec w\cdot \vec z =0$.  This is the way 
that the repeating  IRs in $C_6$ spanned by
$(\vec w, \vec z)$ get identified on the 5-(complex)dimensional cone
$\vec w\cdot \vec z = 0$. 
Redefine,
\begin{eqnarray}
T + m_3 = U,  \qquad  T + n_3 = V 
\label{eq:3.14}		
\end{eqnarray}	
where, $U$ and  $V$ are non-negative integers.
In terms of these variables (eq:~\ref{eq:3.13}) is, 
\begin{eqnarray}
R + U = M \qquad S + V = N   
\label{eq:3.15}	
\end{eqnarray}		
	Also, from (eq:~\ref{eq:3.10})
\begin{eqnarray}
P + Q = R + S
\label{eq:3.16}
\end{eqnarray}		

	We started with freely ranging non-negative integers $m_i$ and 
$n_i$, subject only to the constraints (eq:~\ref{eq:3.2}).  This 
translates to free non-negative integers $P,Q,R,S,U$, and $V$ subject to 
the constraints (eq:~\ref{eq:3.15}) and (eq:~\ref{eq:3.16}).

We have finally arrived at the following explicit and convenient 
realization of the (unnormalized) Gelfand-Zetlin basis of an arbitrary 
IR of $SU(3)$:  Extract the coefficients of various monomials $p^Pq^Q$ in,
\begin{eqnarray}
(pz_1 + qz_2)^R (pw_2 - qw_1)^S z^U_3 w^V_3 
\label{eq:3.17}
\end{eqnarray}
where $w_3$ is given by (eq:~\ref{eq:3.5}). We will denote the space 
spanned by these basis vectors by ${\cal S}(M, N)$.

	In the quark model, the basis within each IR is labeled by the 
quantum numbers $I$ (total isospin), $I_3$ ( the third component of 
isospin) and $Y$ (the hypercharge) (or equivalently strangeness).  These are 
related to our labels $(P Q R S U V)$ as follows:

\begin{eqnarray}
2I = P + Q = R + S,  \qquad  2I_3  =  P - Q, \nonumber\\
Y = \frac{\displaystyle 1}{\displaystyle 3}(M-N) + V - U  \equiv  
\frac{\displaystyle 2}{\displaystyle 3}(N-M) - (S - R)
\label{eq:3.18}				
\end{eqnarray}

	Our labels are better because allowed quantum numbers within 
each IR $(M, N)$ can be read off easily: $R$ takes all values from $0$ 
to $M$.  $S$ takes all values from $0$ to $N$. For a given $R$ and $S$, 
$Q$ takes all values from $0$ to $R + S$.
\section{The generating function}
	Our representation (eq:~\ref{eq:3.17}) of the basis vectors of 
the IRs of $SU(3)$ is not
as simple as the monomial basis of Bargmann for $SU(3)$.  We need to use
specific polynomials.  Inspite of this it has enough features of the 
Bargmann's basis as to be useful for general calculations.  We 
demonstrate this in the following sections.

	The first important feature of our basis is the following.  By 
allowing for all possible non-negative integral values for our labels 
$(P Q R S U V)$ subject to the constraint (eq:~\ref{eq:3.16}), the basis 
for every IR is realized and moreover realized once only.  
Further there are homogeneity  restrictions on
the $\vec{z}$ and $\vec{w}$ variables.  Even though we are forced to use
polynomials instead of the monomials for the basis, the polynomials 
needed can be obtained as the coefficients of a monomial $p^Pq^Q$.  As a 
consequence of all this a simple generating function can be used to 
easily and uniquely generate all unnormalized basis states.
\begin{eqnarray}
{\mbox{\large\bf g}} (p,q,r,s,u,v) = exp(r(pz_1+qz_2) 
+ s(pw_2 - qw_1) + uz_3+vw_3)
\label{eq:3.4.1}
\end{eqnarray}
The coefficient of the ${\mbox{\it monomial}}$,
\begin{eqnarray}
p^P q^Q r^R s^S u^U v^V                                  
\label{eq:3.4.2}
\end{eqnarray}
generates the ${\mbox{\it unnormalized}}$ Gelfand-Zetlin basis denoted by
$\vert P,Q,R,S,U,V)$. Thus,
\begin{eqnarray}
{\mbox{\large\bf g}} = \sum_{P,Q,R,S,U,V} p^Pq^Qr^Rs^Su^Uv^V 
\vert{PQRSUV}) 	\label{eq:3.4.3}
\end{eqnarray}
Note that the constraint (eq:~\ref{eq:3.16}), $P+Q=R+S$, is automatically 
satisfied in the Taylor expansion of (eq:~\ref{eq:3.4.1}).  We refer to 
the variables $p$,$q$,$r$,$s$,$u$
and $v$ as the sources and collectively denote them by $j$.  Similarly, we
refer to the labels $P, Q, R, S, U$ and  $V$ as the quantum numbers and 
collectively denote them by $E$.
	By using the generating function (eq:~\ref{eq:3.4.1}) we have 
come even closer to the Bargmann's techniques for $SU(2)$.

In order to calculate the normalizations of our un-normalized basis 
vectors, we have to first obtain (as in sec.2) the
representation of the generators as differential operators on $\cal G$. 
 An infinitesimal $SU(3)$ matrix may be parameterized as follows:
\begin{eqnarray}
{\cal U} \, \sim \, 1 + i(\epsilon(\pi^0)\pi^0 + \epsilon(\eta)\eta
+\epsilon(\pi^+)\pi^+ + \epsilon(\pi^-) \pi^- +\epsilon(K^+)K^+\nonumber \\
+ \epsilon(K^-)K^- +\epsilon(K^0)K^0 +\epsilon (\overline{K^0})
\overline{K^0} ) \label{eq:3.4.4}
\end{eqnarray}
where the (unnormalized) generators are,
\begin{eqnarray*}
\pi^0 = \pmatrix{1 & 0 & 0 \cr 0 & -1 & 0 \cr 0 & 0 & 0}; \quad 
\eta = \pmatrix{1
& 0 & 0 \cr 0 & 1 & 0 \cr 0 & 0 & -2};                            
\quad \pi^- = \pmatrix{0 & 1 & 0 \cr 0 & 0 & 0 \cr 0 & 0 & 0} 
\end{eqnarray*}
\begin{eqnarray*}
\pi^+ = \pmatrix{0 & 0 & 0 \cr 1 & 0 & 0 \cr 0 & 0 & 0};\quad K^- 
= \pmatrix{0 & 0 & 1 \cr 0 & 0 & 0
\cr 0 & 0 & 0}; \quad K^+ = \pmatrix{0 & 0 & 0 \cr 0 & 0 & 0 \cr 1 & 0 & 0}
\end{eqnarray*}
\begin{eqnarray}
K^0 = \pmatrix{0 & 0 & 0 \cr 0 & 0 & 1 \cr 0 & 0 & 0}; 
\quad \overline{K^0} =
\pmatrix{0 & 0 & 0 \cr 0 & 0 & 0 \cr 0 & 1 & 0}                 
\label{eq:3.4.5}
\end{eqnarray}
For unitarity, we have, 
\begin{eqnarray*}
\epsilon(\pi^0)^* = \epsilon(\pi^0); \quad \epsilon(\eta)^* 
= \epsilon(\eta);
\end{eqnarray*}
\begin{eqnarray*}
\epsilon(\pi^+)^* = \epsilon(\pi^-); \quad \epsilon(K^+)^* = \epsilon(K^-);
\end{eqnarray*}
\begin{eqnarray}
\epsilon(K^0)^* = \epsilon(\bar{K}^0)  
\label{eq:3.4.6}
\end{eqnarray}
Under an $SU(3)$ transformation,
\begin{eqnarray*}
\pmatrix{z_1 \cr z_2 \cr z_3} \rightarrow {\cal U}\pmatrix{z_1 
\cr z_2 \cr z_3}
\end{eqnarray*}
\begin{eqnarray}                                                                
\pmatrix{w_1   & w_2   & w_3} \rightarrow  \pmatrix{w_1   & w_2   & w_3}
{\cal U}^{\dagger}           
\label{eq:3.4.7}
\end{eqnarray}
where ${\cal U}\in SU(3)$.
	This transformation is true even when $w_3$ is eliminated in 
favor of the
other variables since the constraint (eq:~\ref{eq:3.3}) is itself 
invariant under $SU(3)$.  In order to obtain the transformation of 
{\mbox{\large{\bf g}}}, it is convenient to write the exponent in 
(eq:~\ref{eq:3.4.1}) as follows:
\begin{eqnarray}
\pmatrix{r_p & r_q & u }. \pmatrix{z_1 \cr z_2 \cr z_3} + 
\pmatrix{w_1 & w_2 &
w_3 }. \pmatrix{s_q \cr s_p \cr v} 
\label{eq:3.4.8}
\end{eqnarray}
where,
\begin{eqnarray}
r_p = rp, \quad r_q = rq, \quad s_p = sp, \quad s_q = -sq   
\label{eq:3.4.9}
\end{eqnarray}
	It is convenient to regard $r_p$, $r_q$, $s_p$, and $s_q$ as 
independent variables and not related by (eq:~\ref{eq:3.4.9}).  Only at 
the end of the calculations we may set the values (eq:~\ref{eq:3.4.9}) 
and generate the required basis vectors.  We will refer to this operation 
as '{\it{going on shell}}'.  Thus we define the {\it{generalized 
generating function}}:
\begin{eqnarray}
\em{{\cal G}}(r_p, r_q, u, s_q, s_p, v) = exp(\pmatrix{r_p
& r_q & u}\pmatrix{z_1 \cr z_2 \cr z_3} + \pmatrix{w_1 & w_2 & w_3 }.
\pmatrix{s_q \cr s_p \cr v})   
\label{eq:3.4.10}
\end{eqnarray}
Its transformation is,
\begin{eqnarray}
\em{{\cal G}}(r_p, r_q, u, s_q, s_p, v) \rightarrow
\em{{\cal G}}(\pmatrix{r_p & r_q & u}{\cal U}, {\cal U}^{\dagger} 
\pmatrix{s_q \cr s_p \cr v})
\label{eq:3.4.11}
\end{eqnarray}
Substitute (eq:~\ref{eq:3.4.4}) and collect the coefficients of 
$\epsilon(\pi^+)$ etc after using (eq:~\ref{eq:3.4.6}).  We get the 
following representation of the generators:
\begin{eqnarray*}
\hat{\pi}^0 = r_p \frac{\partial}{\partial r_p} - r_q 
\frac{\partial}{\partial r_q} - s_q \frac{\partial}{\partial s_q} 
+ s_p \frac{\partial}{\partial s_p}
\end{eqnarray*}
\begin{eqnarray*}
\hat{\pi}^- = r_p \frac{\partial}{\partial r_q} - s_p \frac{\partial}
{\partial s_q}
\end{eqnarray*}
\begin{eqnarray*}
\hat{\pi}^+ = r_q \frac{\partial}{\partial r_p} - 
s_q \frac{\partial}{\partial s_p}
\end{eqnarray*}
\begin{eqnarray*}
\hat{K}^- = r_p \frac{\partial}{\partial u} - v \frac{\partial}{\partial
s_q}
\end{eqnarray*}
\begin{eqnarray*}
\hat{K}^+ = u \frac{\partial}{\partial r_p} - s_q \frac{\partial}{\partial
v}
\end{eqnarray*}
\begin{eqnarray*}
\hat{K^0} = r_q \frac{\partial}{\partial u} - v \frac{\partial}{\partial
s_p}
\end{eqnarray*}
\begin{eqnarray*}
\hat{\overline{K^0}} = u \frac{\partial}{\partial r_q} - s_p 
\frac{\partial}{\partial v}
\end{eqnarray*}
\begin{eqnarray}
\hat\eta = r_p\frac{\partial}{\partial r_p} + r_q\frac{\partial}
{\partial r_q} - 2u\frac{\partial}{\partial u}-s_p\frac{\partial}
{\partial s_p}-s_q\frac{\partial}{\partial s_q}+2v\frac{\partial}
{\partial v}                
\label{eq:3.4.12}
\end{eqnarray}
	For $\hat{\pi}^0$, $\hat{\pi}^\pm$, and $\hat\eta$ we may use 
(eq:~\ref{eq:3.4.9}) and get the following expressions:
\begin{eqnarray*}
\hat\pi^0 = p\frac{\partial}{\partial p} - q\frac{\partial}{\partial q}
\end{eqnarray*}
\begin{eqnarray*}
\hat\pi^- = p\frac{\partial}{\partial q}
\end{eqnarray*}
\begin{eqnarray*}
\hat\pi^+ = q\frac{\partial}{\partial p}
\end{eqnarray*}
\begin{eqnarray}
\hat \eta = r\frac{\partial}{\partial r} - s\frac{\partial}{\partial s}
-2u\frac{\partial}{\partial u} + 2v\frac{\partial}{\partial v}
\label{eq:3.4.13}
\end{eqnarray}
	However, in order to represent the other generators as differential
operators, we need to regard $r_p$, $r_q$, $s_p$ and $s_q$ as independent
variables.  As a result we face the following problem.  It is not easy to
calculate the matrix elements,
\begin{eqnarray}
(P' Q' R' S' U' V'\vert T \vert P Q R S U V)
\label{eq:3.4.14}
\end{eqnarray}
of these generators, which are needed to evaluate the normalizations (see
Sec 2).  Consider for example the action of $\hat{K}^-$ on {\large\bf g}:
\begin{eqnarray}
\hat{K}^- {\mbox{\large\bf g}} = (rpz_3 - vw_1){\mbox{\large\bf g}}
\label{eq:3.4.15}
\end{eqnarray}
	We need to re-express the effect of $z_3$ or $w_1$ multiplying a 
basis state (eq:~\ref{eq:3.17}) as a linear combination of such states.  
But this is not easy.
	
	This is another stumbling block compared to situation in $SU(2)$.  We
devise the technique to overcome this problem  in the next section.

\section{An auxiliary Gaussian measure}
	An important reason for the efficiency of Bargmann's techniques for
$SU(2)$ is the Gaussian measure, using which calculations can be performed
explicitly and easily.  It is obtained as the measure with respect to which
the Gelfand-Zetlin vectors form an orthonormal set and the representation
matrices are unitary.  We have discussed this in Sec.2.

	We have a model space using five complex variables 
$(z_1, z_2, z_3, w_1, w_2)$.  The measure in this space with respect 
to  which properly normalized
basis vectors form an orthonormal set exists, in principle.  Gelfand et.al
\cite{BIN} have obtained this measure as the solution of a differential 
equation in a related context.  Unfortunately, this measure does not 
have the simplicity of the Gaussian measure for $SU(2)$.  It appears that 
using it as a calculational tool to obtain general formulae is quite 
remote.  This is yet another 
stumbling block in extending the Bargmann techniques to $SU(3)$.

	We evade this problem in the following way.  We construct an 
auxiliary measure which is amenable to computations by relaxing the 
condition that it gives correct {\it {normalizations}} of the Gelfand-
Zetlin states.  We only require that the basis states do form a 
{\it {orthogonal}} set with respect to the measure.  This is in fact 
sufficient for calculations of formulae.  The correct normalization of 
the basis states (which gives a unitary representation) is itself 
computed using the same measure.

	Our condition on the measure may be expressed in terms of the 
generating function as follows.  Define the inner product, 
\begin{eqnarray}
({\mbox{\large\bf g}'}, {\mbox{\large\bf g}}) &=& \int d\mu 
\quad exp\overline{\left(r'(p'z_1+q'z_2)+s'(p'w_2-q'w_1)+u'z_3
+v'w_3\right)}\nonumber\\ 
& & exp\left(r(pz_1+qz_2)+s(pw_2-qw_1)+uz_3+vw_3\right)                    
\label{eq:3.5.1}
\end{eqnarray}
between generating functions for different sets of arguments.  (The over
line in the first exponential means complex conjugation of the expression 
under it).  The integration is over the variables 
$(z_1, z_2, z_3, w_1, w_2)$, $w_3$ being expressed in terms of these 
other variables.  We want this inner product to be
of the following type.
\begin{eqnarray}
({{\mbox{\large\bf g}}'}, {\mbox{\large\bf g}}) = F(\bar{p}'p, 
\bar{q}'q, \bar{r}'r, \bar{s}'s, \bar{u}'u, \bar{v}'v)
\label{eq:3.5.2}
\end{eqnarray}
where the function $F$ has a Taylor expansion in its arguments about the 
origin with every coefficient positive definite.  Such a form implies 
that the inner product is zero whenever the powers of the variables 
$(p, q, r, s, u, v)$do not match the powers of the corresponding 
variables $(p', q', r', s', u', v')$,
i.e., Gelfand-Zetlin basis vectors are mutually orthogonal.  Moreover the
square of the norm given by the corresponding coefficient in the Taylor
expansion is positive definite.

	Our measure is closely related to Bargmann's.  We have two 
doublets $(z_1, z_2)$ and $(w_1, w_2)$ of $SU(2)$ and the coupled basis 
built using them.  We know that coupled basis is obtained by an unitary 
transformation of the direct product basis.
This means that Bargmann's measure for these two doublets ensures
orthogonality of our $SU(3)$ basis vectors in so far as the
$(z_1,z_2,w_1,w_2)$ variables are concerned.  To be explicit we consider,
\begin{eqnarray}
F \qquad = \qquad \int
\frac{d^2z_1}{\pi} \frac{d^2z_2}{\pi} \frac{d^2w_1}{\pi}
\frac{d^2w_2}{\pi}exp(-\bar z_1 z_1-\bar z_2 z_2-\bar w_1 w_1
-\bar w_2 w_2)\nonumber\\
\times exp\left(\overline{(r'(p'z_1+q'z_2)+s'(p'w_2-q'w_1)- 
\frac{v'}{z_3}(z_1w_1+z_2w_2)+u'z_3}\right)\nonumber \\
\times exp\left({r(pz_1 + qz_2) + s(pw_2 - qw_1) - \frac{v}{z_3}
(z_1w_1 + z_2w_2) +
uz_3}\right)                                                      
\end{eqnarray}
This can be evaluated easily.  Write the exponent as, 
\begin{eqnarray}
-\pmatrix{(\bar{z}_1 & w_1} \pmatrix{1 & \overline{(\frac{v'}{z_3})} 
\cr \frac{v}{z_3}
& 1} \pmatrix{z_1 \cr \bar{w}_1} -
 \pmatrix{(\bar{z}_2 & w_2} \pmatrix{1 & 
\overline{(\frac{v'}{z_3})} \cr \frac{v}{z_3} & 1} \pmatrix{z_2 \cr 
\bar{w}_2}\nonumber\\
+ \pmatrix{rp & - \overline{(s'q'})} \pmatrix{z_1 \cr \bar{w}_1} 
+ \pmatrix{\bar{z}_1 & w_1} \pmatrix{\overline{r'p'} \cr -sq} +
\pmatrix{rq & (\overline{s'p'})} \pmatrix{z_2 \cr \bar{w}_2}\nonumber\\
+ \pmatrix{\bar{z}_2 & w_2} \pmatrix{\overline{(r'q')} \cr sp} 
+ \overline{(u'{z_3})} + uz_3  \
\label{eq:3.5.4}
\end{eqnarray}
Use the formula,
\begin{eqnarray}
\prod^n_{i=1}\int{\frac{d^2z_i}{\pi}}exp(-\bar{z}^T X z + A^Tz + \bar{z}^T
\bar{B}) = (detX)^{-1} exp(A^T X^{-1}\bar{B})
\label{eq:3.5.5}
\end{eqnarray}
which is valid whenever the hermitian part of $X$ is positive definite. 
 Here $z$ is the column vector of the complex variables 
$(z_1, z_2, ...,z_n)$.  We get,
\begin{eqnarray}
F = (1 - \frac{\bar{v}'v}{\vert{z_3}\vert^2})^{-2} exp[((1 - {\bar{v}'v}
/{\vert{z_3}\vert^2)^{-1}} (\pmatrix{rp & -\bar{s}'\bar{q}'}\pmatrix
{1 & -{\bar{v}'}/{\bar{z_3}} \cr -{v}/{z_3} & 1}\nonumber \\
\times \pmatrix{\bar{r}'\bar{p}' \cr -sq} +\pmatrix{rq & \bar{s}'\bar{p}'}
\pmatrix{1 & -{\bar{v}'}/{\bar{z}_3} \cr -{v}/{z_3} & 1}\pmatrix
{\bar{r}'\bar{q}' \cr sp}) +\bar{u}'\bar{z}_3+uz_3]  
\end{eqnarray}
Thus,
\begin{eqnarray}
F = (1 -
\frac{\bar{v}'v}{\vert{z_3}\vert^2})^{-2} exp(\frac{(\bar{r}'r +
\bar{s}'s)(\bar{p}'p + \bar{q}'q)}{(1 -
{\bar{v}'v}/{\vert{z_3}\vert^2})}+\bar{u}'\bar{z}_3+uz_3) 
\label{eq:3.5.6}
\end{eqnarray}
	Note that this has the features (eq:~\ref{eq:3.5.2}) we required 
for the inner product, so far as the variables $p, q, r, s$ and $v$ are 
concerned.  This was to be
expected, because the coupled basis constructed out of two doublets
$\pmatrix{z_1 & z_2}$ and $\pmatrix{w_1 & w_2}$ is an orthogonal set with
respect to Bargmann's measure.

	We have to now propose a workable measure for integration over 
the $z_3$ variable.  Note that $\vert{z_3}\vert^2$ appears in the 
denominators in (eq:~\ref{eq:3.5.6}).
 This suggests that it is best to set,
\begin{eqnarray}
z_3 = e^{i\theta}                                               
\label{eq:3.5.7}
\end{eqnarray}
so that $\vert{z_3}\vert^2 = 1$.
	This means that our realization of the basis vectors is in terms 
of four complex variables $z_1, z_2, w_1, w_2$ and a phase variable 
$e^{i\theta}$.  Setting the constraint (eq:~\ref{eq:3.5.7}) is no problem, 
because we have used only the variable $z_3$ and not $\bar{z}_3$ in our 
basis vectors.  (See Appendix A for a more detailed discussion.) Our 
requirement on the measure for the $\theta$ variable is that (i)it is 
simple and (ii)we get a function of only the combination $\bar{u}'{u}$.  
>From (eq:~\ref{eq:3.5.6}) we see that it
is sufficient to average over $\theta$.  Thus the measure we use is,
\begin{eqnarray}
({\mbox{\large\bf g}}', {\mbox{\large\bf g}}) = 
\int^{+\pi}_{-\pi}\frac{d\theta}{2\pi}\int
\frac{d^2z_1}{\pi} \frac{d^2z_2}{\pi} \frac{d^2w_1}{\pi}
\frac{d^2w_2}{\pi}\bar{\mbox{\large\bf g}}'{\mbox{\large\bf g}}                \label{eq:3.5.8}
\end{eqnarray}
The inner product between two generating functions is,
\begin{eqnarray}
({\mbox{\large\bf g}}', {\mbox{\large\bf g}}) 
= (1-\bar{v}'v)^{-2}(\sum^\infty_{n=0}\frac{(\bar{u}'{u})^n}{(n!)^2})
exp[(1-\bar{v}'v)^{-1}(\bar{p}'p + \bar{q}'q)(\bar{r}'r + \bar{s}'s)]
\label{eq:3.5.9}
\end{eqnarray}
Notice that the coefficients of the Taylor expansion are all positive 
definite.  This is a satisfactory inner product using which we may do 
computations explicitly.

	For our calculations, we need the inner product between any two 
generalized generating functions (eq:~\ref{eq:3.4.10}).  In place of 
(eq:~\ref{eq:3.5.6}), we get,
\begin{eqnarray}
({\cal G}', {\cal G}) = \int  \frac{d\theta}{2\pi}(1-\bar{v}'v)^{-2} exp
[(1-\bar{v}'v)^{-1} \pmatrix{r_p & \bar{s_q}'}\pmatrix{1 & -\bar{v}'
e^{i\theta} \cr -{v}e^{-i\theta} & 1}\nonumber\\
\pmatrix{\bar{r_p}' \cr s_q} + \pmatrix{(r_q & \bar{s_p}'}\pmatrix{1 &
-\bar{v}'e^{+i\theta} \cr -{v}e^{-i\theta} & 1}\pmatrix{\bar{r_q}' \cr
s_p}\nonumber\\
 +\bar{u}'e^{-i\theta}+ue^{i\theta}]
\end{eqnarray}
Therefore, 
\begin{eqnarray}
({\cal G}', {\cal G}) =
(1 - \bar{v}'v)^{-2}exp[(1 - \bar{v}'v)^{-1}(\bar{r_p}'r_p 
+ \bar{r_q}'r_q +
\bar{s_p}'s_p + \bar{s_q}'s_q)]\nonumber \\
\times [\sum^{\infty}_{n=0} \frac{1}{(n!)^2}\left (\bar{u}' 
- v\frac{(\bar{r_p}'\bar{s_q}' + \bar{r_q}'\bar{s_p}')}{(1 - \bar{v}'v)}
\right )^n \left ({u} - \bar{v}'\frac{({r_p}{s_q} + {r_q}{s_p})}{(1 
- \bar{v}'v)}\right )^n]
\label{eq:3.5.10}
\end{eqnarray}
\section{Calculation of the normalizations}
We now compute the normalization of our representation 
(eq:~\ref{eq:3.17}) of the (unnormalized) basis vectors.  As discussed in section 2, 
this is obtained from the requirement that the representation matrix be 
unitary in each IR.  Our technique differs from the one discussed in 
section 2 in one crucial respect.

	Let $\vert E)$ denote any  unnormalized basis vector defined by 
the expansion (eq:~\ref{eq:3.4.3}). $E$ stands for the set of quantum 
numbers used in the basis.  For any generator $T$ let,
\begin{eqnarray}
T \vert E) = \sum_{E'} T(E, E') \vert E')
\label{eq:6.1}
\end{eqnarray}
We want to compute $N(E)$ defined by,
\begin{eqnarray}
\vert E) = N^{\frac{1}{2}}(E) \vert E >         
\label{eq:6.2}
\end{eqnarray}
where $\vert E >$ denotes any Gelfand-Zetlin normalized basis vector :
\begin{eqnarray}
< E' \vert E > = \delta_{E'E}            
\label{eq:6.3}
\end{eqnarray}
The representation matrix is unitary when the basis vectors $\vert E >$
are used.  $N(E)$ is obtained by the requirement,
\begin{eqnarray}
< E' \vert T \vert E > = < E \vert T^* \vert E' >^*
\label{eq:6.4}
\end{eqnarray}
for every generator $T$ and for any normalized basis vectors $\vert E >$
and $\vert E' >$.  We have from (eq:~\ref{eq:6.1}) and (eq:~\ref{eq:6.2}),
\begin{eqnarray}
N^{\frac{1}{2}}(E) T\vert E> = \sum_{E'} T(E,E') N^\frac{1}{2}(E')\vert
E'>    
\label{eq:6.5}
\end{eqnarray}
so that,
\begin{eqnarray}
< E' \vert T \vert E > = T(E,E')
\frac{N^\frac{1}{2}(E')}{N^\frac{1}{2}(E)}  
\label{eq:6.6}
\end{eqnarray}
Therefore, (eq:~\ref{eq:6.4}) gives,
\begin{eqnarray}
\left | \frac{N(E)}{N(E')} \right | = \frac{T(E,E')} {(T(E',E))^*}
\label{eq:6.7}
\end{eqnarray}
This means we need to evaluate $T(E,E')$, defined in (eq:~\ref{eq:6.1}).  
For this we use our 'auxiliary' inner product given by (eq:~\ref{eq:3.5.8}).  
We denote this inner product between two vectors 
$\vert 1)$ and $\vert 2)$ by
\begin{eqnarray}
(2 \vert \vert 1) 
\label{eq:6.8}
\end{eqnarray}
to distinguish it from the 'true' inner product given by 
(eq:~\ref{eq:6.3}).  The
Gelfand-Zetlin normalized basis vectors $\vert E >$ do form an orthogonal
set but have a different norm w.r.t. the auxiliary inner product. Therefore,
\begin{eqnarray}
(E' \vert \vert E ) = \delta_{E'E} M(E)
\label{eq:6.9}
\end{eqnarray}
where $M(E)$ is different from $N(E)$ in general.
Using (eq:~\ref{eq:6.9}) in (eq:~\ref{eq:6.1}) we get,
\begin{eqnarray}
(E' \vert \vert ( T \vert E)) = T(E,E') M(E')
\label{eq:6.10}
\end{eqnarray}
Therefore (eq:~\ref{eq:6.7}) gives,
\begin{eqnarray}
\left  |  \frac{N(E)}{N(E')}\right | = \frac{(E'  \vert  \vert  T 
\vert E)}{(E \vert \vert T^*
\vert E')^*} \frac{M(E)}{M(E')}  
\label{eq:6.11}
\end{eqnarray}
Thus we can fix the normalization using an 'auxiliary' inner product which
allows explicit computation, even though it is not the 'true' inner product.

$M(P,Q,R,S,U,V)$ can be read off as the coefficient of the monomial:
\begin{eqnarray*}
(\bar{p'}p)^P(\bar{q'}q)^Q (\bar{r'}r)^R (\bar{s'}s)^S (\bar{u'}u)^U 
(\bar{v'}v)^V
\end{eqnarray*}
in (eq:~\ref{eq:3.5.9}).

We have,
\begin{eqnarray}
({\mbox{\large\bf g}}',{\mbox{\large\bf g}})=\left (\sum^\infty_{2I=0}
\frac{(\bar{p'}p + \bar{q'}q)^{2I}(\bar{r'}r + \bar{s'}s)^{2I}}{(2I)!
(1-\bar{v'}v)^{2I+2}}\right )
\left (\sum^\infty_{V=0}\frac{(\bar{u'}u)^U}{(U!)^2}\right )
\label{eq:6.12}
\end{eqnarray}
Using
\begin{eqnarray}
(x+y)^n = \sum^n_{m=0} {^n}C_m x^m y^{n-m} 
\label{eq:6.13}
\end{eqnarray}
and 
\begin{eqnarray}
\frac {1}{(1-x)^{m+1}} = \sum^{\infty}_{n=0} {^{(n+m)}}C_n x^n 
\label{eq:6.14}
\end{eqnarray}
we get,
\begin{eqnarray}
M(P,Q,R,S,U,V)	= \frac{(V+2I+1)!}{P!Q!R!S!(U!)^2V!(2I+1)}
\label{eq:6.15}
\end{eqnarray}
We have used $2I= P+Q= R+S$.

	We now apply formula (eq:~\ref{eq:6.11}) for each of our 
generators (eq:~\ref{eq:3.4.12}).  The
generators $\hat\pi^0$ and $\hat\eta$ are diagonal in the chosen basis 
and therefore do not lead to any constraints on the relative 
normalizations of the basis vectors.
	Consider $\hat\pi^-$ as given by (eq:~\ref{eq:3.4.13}).
We get,
\begin{eqnarray}
({\mbox{\large\bf g}}',{\hat\pi}^-{\mbox{\large\bf g}})= p  
\frac{\partial}{\partial  q} 
({\mbox{\large\bf g}}',{\mbox{\large\bf g}}) = p\bar{q}'\frac{(\bar{r}'r
+\bar{s}'s)}{(1-v'v)} ({\mbox{\large\bf g}}',{\mbox{\large\bf g}})
\label{eq:6.16}
\end{eqnarray}
on using (eq:~\ref{eq:3.5.9}).
Extracting like powers of the monomials from both sides of 
(eq:~\ref{eq:6.13}), we get
\begin{eqnarray}
(P,Q+1,R,S,U,V) \| \hat\pi^- \vert P+1,Q,R,S,U,V) = M_3(P,Q,R,S,U,V)
\label{eq:6.17}
\end{eqnarray}
as the only non-zero auxiliary matrix element of $\hat\pi^-$.  This is 
expected because $\hat\pi^-$ only lowers $I_3$ value  by 1. (see 
(eq:~\ref{eq:3.18})).
$M_3(P,Q,R,S,U,V)$ is listed in Table 1.
Similarly we get,
\begin{eqnarray}
(P+1,Q,R,S,U,V  \vert  \vert  \hat\pi^+  \vert  P,Q+1,R,S,U,V)  = 
M_3(P,Q,R,S,U,V)
\label{eq:6.18}
\end{eqnarray}

Thus, in the present case,
\begin{eqnarray}
(P,Q+1,R,S,U,V  \vert  \vert \hat\pi^- \vert  P+1,Q,R,S,U,V)\nonumber \\ 
 = (P+1,Q,R,S,U,V \vert \vert\hat\pi^+ \vert  P,Q+1,R,S,U,V)
\label{eq:6.19}
\end{eqnarray}
since $M_3(P,Q,R,S,U,V)$ is real.  
Therefore the auxiliary normalization, also gives the Gelfand-Zetlin
normalizations,  in this case:
\begin{eqnarray}
\left | \frac{N(P+1, Q, R, S, U, V)}{N(P, Q+1, R, S, U, V)} \right |
 = \frac{M(P+1, Q, R, S, U, V)}{M(P,Q+1,R, S, U, V)}
\label{eq:6.20}
 \end{eqnarray}
 The reason for this matching is that for the $SU(2)$ subgroup we are 
using just the Bargmann measure.
	Using Table 1 we get,
\begin{eqnarray}
\left| \frac{N(P+1, Q, R, S, U, V)}{N(P, Q+1, R, S, U, V)} \right| = 
\frac{Q+1}{P+1}
\label{eq:6.21}
\end{eqnarray}
exactly as in the $SU(2)$ case (section 2).
Thus the relative normalizations of basis vectors within an isospin 
multiplet are determined and are same as in the $SU(2)$ case :
\begin{eqnarray}
N(P, Q, R, S, U, V) \sim \frac{1}{P!Q!} 
\label{eq:6.22}
\end{eqnarray}
The dependence on the total isospin $ P+Q = R+S$ as also on quantum 
numbers $R$, $S$, $U$ and $V$ are not determined at this stage.

	We now compute the relative normalizations implied by 
$\hat{K}^\pm$. To calculate, $({\mbox{\large\bf g}}',\hat{K}^-{\mbox
{\large\bf g}})$ we use the generalized partition function:
\begin{eqnarray}
({\mbox{\large\bf g}}', \hat{K}^-{\mbox{\large\bf g}}) = ( r_p 
\frac{\partial}{\partial u} - v{\frac{\partial}{\partial s_q}})
(\cal {G}',\cal {G}) \vert             
\label{eq:6.23}
\end{eqnarray}
where the vertical line at the end of this equation means that after 
applying differential operator on $(\cal {G}',\cal {G})$, we need to 
set the values (eq:~\ref{eq:3.4.9}) for the sources.
For instance,
\begin{eqnarray}
(r_ps_q + r_qs_p){\large |} = 0; \quad  (\bar{r}_p'\bar{s}_q' + 
\bar{r}_q'\bar{s}_p') {\large |}= 0
\label{eq:6.24}  
\end{eqnarray}
We get, 
\begin{eqnarray}
({\mbox{\large\bf g}}', \hat{K}^-{\mbox{\large\bf g}})=\frac{1}
{(1-\bar{v}'v)^2}exp[(1-\bar{v}'v)^{-1}(\bar{p}'p+\bar{q}'q)
(\bar{r}'r + \bar{s}'s)] \nonumber \\
\times \left [rp \sum^\infty_{n=0} \frac {\bar{u'}^{n+1}u^n}{(n+1)!n!}
+\frac{v\bar{s}'\bar{q}'}{(1-\bar{v}'v)} \sum^\infty_{n=0} 
\frac{\bar{u'}^nu^n}{(n!)^2}
+\frac{v\bar{v}'rp}{(1-\bar{v}'v)} \sum^\infty_{n=0}
\frac{\bar{u}'^{n+1}u^n}{(n+1)!n!}\right ]                           
\label{eq:6.25}
\end{eqnarray}
Matching coefficients of like powers we get (Table 1),
\begin{eqnarray}
(P,Q,R,S,U+1,V \vert \vert \hat{K}^- \vert P+1,Q,R+1,S,U,V)\nonumber \\
 = M_5(P,Q,R,S,U,V)   + M_6(P,Q,R,S,U,V-1)                     
\label{eq:6.26}
\end{eqnarray}
\begin{eqnarray}
(P,Q+1,R,S+1,U,V  \vert  \vert \hat{K}^- \vert  P,Q,R,S,U,V+1)\nonumber \\
= M_1(P,Q,R,S,U,V)
\label{eq:6.27}
\end{eqnarray}
These non-zero matrix elements are as expected for $I=\frac{1}{2},
I_3=-\frac{1}{2}, Y=+1$ quantum numbers for $\hat{K}^-$.
Similarly,
\begin{eqnarray}
(P+1,Q,R+1,S,U,V) \vert \vert \hat{K}^+ \vert P,Q,R,S,U+1,V)\nonumber \\
 = M_1(P,Q,R,S,U,V)
\label{eq:6.28}
\end{eqnarray}
\begin{eqnarray}
(P,Q,R,S,U,V+1 \vert \vert \hat{K}^+ \vert P,Q+1,R,S+1,U,V) =~~~~~~~~~~
  \nonumber \\
M_6(P,Q,R,S,U\!-\! 1,V)\!+\!2M_1(P,Q,R,S,U,V)+M_4(P,Q,R,S,U,V)
\label{eq:6.29}
\end{eqnarray}
Using (eq:~\ref{eq:6.12}) we get the following constraints on relative 
normalizations from (eq:~\ref{eq:6.24})-(eq:~\ref{eq:6.27}) :
\begin{eqnarray}
\left |\frac{N(P+1,Q,R+1,S,U,V)}{N(P,Q,R,S,U+1,V)}\right | 
=(V+2I+2)\frac{(U+1)}{(P+1)(R+1)}
\frac{(2I+1)}{(2I+2)}                                        
\label{eq:6.30}
\end{eqnarray}
\begin{eqnarray}
\left |\frac{N(P,Q+1,R,S+1,U,V)}{N(P,Q,R,S,U,V+1)}\right | =(U+2I+2) 
\frac{(V+1)}{(Q+1)(S+1)}
\frac{(2I+1)}{(2I+2)}                                       
\label{eq:6.31}
\end{eqnarray}
	A solution for (eq:~\ref{eq:6.28}) and (eq:~\ref{eq:6.29}) is, 
\begin{eqnarray}
N(P,Q,R,S,U,V)=\frac{(U+2I+1)!     (V+2I+1)!}{P!Q!R!S!U!V!(2I+1)}	 
\label{eq:6.32}
\end{eqnarray}

We now consider the non-uniqueness in this solution.  The quantum numbers,
$Q$,$S$,$V$, $P+U$, and $R+U$ do not change between the numerator and 
the denominator of (eq:~\ref{eq:6.30}).  Therefore dependence of 
$N(P,Q,R,S,U,V)$ on these quantum numbers are not fixed by (eq:
~\ref{eq:6.30}).  However, (eq:~\ref{eq:6.31}) serves to fix the 
dependence on $Q$,$S$,$V$.  Therefore the only ambiguity is in 
dependence of the combinations $P+U$ and $R+U$.  We may hope that this 
ambiguity is removed by the constraints coming from the remaining 
generators $\hat{{K}^0}$ and $\hat{\bar{K^0}}$.  The non-zero matrix 
elements of these generators in the unnormalized basis are:
\begin{eqnarray}
(P,Q,R,S,U+1,V \vert \vert \hat{K}^0 \vert P,Q+1,R+1,S,U,V)\nonumber \\
 =M_5(P,Q,R,S,U,V) +M_6(P,Q,R,S,U,V-1)                                            
\label{eq:6.33}
\end{eqnarray}
\begin{eqnarray}
(P+1,Q,R,S+1,U,V \vert \vert \hat{{K}^o} \vert P,Q,R,S,U,V+1)\nonumber \\
= - M_1(P,Q,R,S,U,V)
\label{eq:6.34}
\end{eqnarray}
\begin{eqnarray}
(P,Q+1,R+1,S,U,V \vert \vert {\hat{\bar {K^o}}} \vert P,Q,R,S,U+1,V)
\nonumber \\
= M_1(P,Q,R,S,U,V)                                                
\label{eq:6.35}
\end{eqnarray}
\begin{eqnarray}
(P,Q,R,S,U,V+1) \vert \vert {\hat{\bar {K^0}}} \vert P+1,Q,R,S+1,U,V)
\nonumber \\
 =\! -\!M_6(\!P,Q,R,S,U\!-\!1,V\!)\!-\!2\!M_1(\!P,Q,R,S,U,V\!)\!
-\!M_4(\!P,Q,R,S,U,V\!) 
\label{eq:6.36}
\end{eqnarray}
This gives the constraints,
\begin{eqnarray}
\left |\frac {N(P,Q+1,R+1,S,U,V)}{N(P,Q,R,S,U+1,V)}\right | 
= (V+2I+2) \frac{(U+1)}{(Q+1)(R+1)} \frac{(2I+1)}{(2I+2)}                                           
\label{eq:6.37}
\end{eqnarray}
\begin{eqnarray} 
\left |\frac{N(P+1,Q,R,S+1,U,V)}{N(P,Q,R,S,U,V+1)}\right |=(U+2I+2)
\frac{(V+1)}{(P+1)(S+1)}\frac{(2I+1)}{(2I+2)}                  
\label{eq:6.38}
\end{eqnarray} 
As expected, these constraints fix the dependence on $P+U$ and $R+U$.
	Thus the normalization factor of our unnormalized basis states 
is uniquely fixed by (eq:~\ref{eq:6.32}).  It is worth noting that the 
matrix elements (eq:~\ref{eq:6.27}), (eq:~\ref{eq:6.29}) and (eq:~\ref
{eq:6.35}) are all equal and differ  only in sign from (eq:~\ref{eq:6.34}).  
Similarly, (eq:~\ref{eq:6.26}) and (eq:~\ref{eq:6.33}) are equal, whereas, 
(eq:~\ref{eq:6.29}) only differs in sign from (eq:~\ref{eq:6.36}).  As a 
consequence, there is a certain symmetry in the constraints (eq:~\ref
{eq:6.30}), (eq:~\ref{eq:6.31}), (eq:~\ref{eq:6.37}) and (eq:~\ref
{eq:6.38}) of the normalizations.  These relations between matrix 
elements   with  respect  to  auxiliary  inner  product   between 
unnormalized basis states is a  consequence 
of  our choices of the inner product and basis states.   They  do 
not seem to have any group theoretic reason.  Anyway we are  only 
interested in matrix elements between normalized states.
\newline
{\small
\begin{table}
\centering
\caption{Normalization Constants.}
$$
\begin{array}{|l|l|l|}
\hline
M(P,Q,R,S,U,V) & \frac{exp[(1-\bar{v}'v)^{-1}(\bar
{p}'p+\bar{q}'q)(\bar{r}'r+\bar{s}'s)]}{(1-\bar{v}'v)^{2}}\sum_n
\frac{(\bar{u}'u)^n}{(n!)^2}
& \frac{(2I+1+V)!}{P!Q!R!S!U!V!}\frac{1}{U!(2I+1)}\\ \hline 
M_1(P,Q,R,S,U,V) & {\frac{exp[(1-\bar{v}'v)^{-1}(\bar
{p}'p+\bar{q}'q)(\bar{r}'r+\bar{s}'s)]}{(1-\bar{v}'v)^{3}}\sum_n\frac
{(\bar{u}'u)^n}{(n!)^2}}
& \frac{(2I+2+V)!}{P!Q!R!S!U!V!}\frac{1}{U!(2I+1)(2I+2)}\\ \hline
M_2(P,Q,R,S,U,V) & \frac{exp[(1-\bar{v}'v)^{-1}(\bar
{p}'p+\bar{q}'q)(\bar{r}'r+\bar{s}'s)]}{(1-\bar{v}'v)^{4}}\sum_n\frac
{(\bar{u}'u)^n}{(n!)^2}
& \frac{(2I+3+V)!}{P!Q!R!S!U!V!}\frac{1}{U!(2I+1)(2I+2)(2I+3)}\\ \hline
M_3(P,Q,R,S,U,V) & \frac{exp[(1-\bar{v}'v)^{-1} (\bar
{p}'p+\bar{q}'q)(\bar{r}'r+\bar{s}'s)]}{(1-\bar{v}'v)^{3}} & 
{\mbox{Not needed}}\\
                 & \times (\bar{r}'r+\bar{s}'s)
\sum_n\frac{(\bar{u}'u)^n}{(n!)^2} & \\ \hline
M_4(P,Q,R,S,U,V) & \frac{exp[(1-\bar{v}'v)^{-1}(\bar
{p}'p+\bar{q}'q)(\bar{r}'r+\bar{s}'s)]}{(1-\bar{v}'v)^{4}} & 
\frac{(2I+2+V)!}{P!Q!R!S!U!V!}\frac{2I}{U!(2I+1)(2I+2)}\\
& \times (\bar
{p}'p+\bar{q}'q)(\bar{r}'r+\bar{s}'s) \sum_n\frac{(\bar{u}'u)^n}{(n!)^2}
& \\ \hline
M_5(P,Q,R,S,U,V) & \frac{exp[(1-\bar{v}'v)^{-1}(\bar
{p}'p+\bar{q}'q)(\bar{r}'r+\bar{s}'s)]}{(1-\bar{v}'v)^{2}}\sum_n
\frac{(\bar{u}'u)^n}{(n+1)!n!}
& \frac{(2I+1+V)!}{P!Q!R!S!U!V!}\frac{1}{(U+1)!(2I+1)}\\ \hline
M_6(P,Q,R,S,U,V) & \frac{exp[(1-\bar{v}'v)^{-1}(\bar
{p}'p+\bar{q}'q)(\bar{r}'r+\bar{s}'s)]}{(1-\bar{v}'v)^{3}}\sum_n\frac
{(\bar{u}'u)^n}{(n+1)!n!}
& \frac{(2I+2+V)!}{P!Q!R!S!U!V!}\frac{1}{(U+1)!(2I+1)(2I+2)}\cr
\hline
\end{array}
$$
\end{table}
}
\section{The 3-G symbols}
In  this  section,  we  mostly  review  well  -  known   material 
\cite{BAO,ZDP,MM1,LE,RM} in order to fix  our 
notations and  for 
logical  continuity.   In  addition we  emphasize  the   relation 
between  the multiplicity in the Clebsch - Gordan series and  the 
distinct invariants that can be constructed out of three IRs.  
Consider a compact group G.  We denote the basis of states of unitary
irreducible representations by
\begin{eqnarray}
\vert^\lambda_\alpha \rangle 
\label{eq:7.1}
\end{eqnarray}
where $\lambda$ labels the IRs and $\alpha$ labels the basis for the IRs.  
For $SU(3)$, for instance, we may use the ordered pair $(M,N)$ for 
$\lambda$ and $(I,I_3,Y)$ for $\alpha$.  Consider the direct product of 
two IRs, $\lambda$ and $\lambda'$.  This can be completely reduced to the 
IRs of the group.  In general, same IR $\lambda''$ may appear more than 
once in the decomposition.  Therefore we need extra labels for the IRs 
of the decomposition.  We denote these additional labels collectively by 
$k$.  (We discuss this in detail \cite{AGMS,GS} for $SU(3)$ in sec.8).  
Thus we may write,
\begin{eqnarray}
\vert^{\lambda"}_{\alpha"} \rangle^k =
\sum_{\alpha,\alpha'} \vert^\lambda_\alpha
\rangle\vert^{\lambda'}_{\alpha'}\rangle\langle^{\lambda
\lambda'}_{\alpha\alpha'}\vert^{\lambda"}_{\alpha"}\rangle^k
\label{eq:7.2}
\end{eqnarray}
where $\vert^{\lambda"}_{\alpha"}\rangle^k$ are the basis vectors of the
repeating IR $\lambda"$ and where the repetitions are labeled by $k$.  
The coefficients in this expansion,
\begin{eqnarray}
\langle^{\lambda \lambda'}_{\alpha \alpha'}\vert^{\lambda"}_{\alpha"}
\rangle^k \label{eq:7.3}
\end{eqnarray}
are the Wigner-Clebsch-Gordan coefficients of $G$.
(eq:~\ref{eq:7.2}) provides an unitary transformation from the basis of 
the IRs of the decomposition to the direct product basis.  Inverse 
transformation may be written,  
\begin{eqnarray}
\vert^\lambda_\alpha\rangle\vert^{\lambda'}_{\alpha'}\rangle = 
\sum_{\alpha",k} \vert^{\lambda"}_{\alpha"}\rangle^k \, {^k}\langle^
{\lambda"}_{\alpha"}\vert^{\lambda\lambda'}_{\alpha\alpha'}\rangle
\label{eq:7.4}
\end{eqnarray}
We get,
\begin{eqnarray}
\sum_{\lambda",k,\alpha"}\langle^{\lambda
\lambda'}_{\alpha\alpha'}\vert^{\lambda"}_{\alpha"}{\rangle^k} \, 
{^k}\langle^{\lambda" }_{\alpha"}\vert^{\lambda\lambda'}_{\beta\beta'}
\rangle = \delta_{\alpha\beta}\delta_{\alpha'\beta'}  \label{eq:7.5}
 \end{eqnarray}
\begin{eqnarray}
\sum_{\alpha,\alpha'} {^k}\langle^{\lambda"
}_{\alpha"}\vert^{\lambda\lambda'}_{\alpha\alpha'}\rangle\langle^{\lambda
\lambda'}_{\alpha\alpha'}\vert^{\lambda^{'''}}_{\alpha^{'''}}\rangle^{k'} =
\delta_{kk'}\delta_{\lambda"\lambda{'''}}\delta_{\alpha"\alpha{'''}}
\label{eq:7.6}
\end{eqnarray}
Let $D(g)^\lambda$ denote the unitary representation matrix of an element
$g\in G$ in the IR $\lambda$.  Thus under the group action, 
\begin{eqnarray}
g:\vert^\lambda_\alpha \rangle \rightarrow \sum_\beta
D^\lambda_{\alpha\beta}(g) \vert^\lambda_\beta\rangle
\label{eq:7.7}
\end{eqnarray}
The matrices $\overline {D}^{\lambda}(g)$ whose elements are complex 
conjugates of $D^{\lambda}(g)$,
\begin{eqnarray}
(\overline {D}(g))_{\alpha \beta} = (D^{\lambda}(g))^*_{\alpha \beta}
\label{eq:7.8}
\end{eqnarray}
also provide an irreducible representation of the group called the
representation $\bar \lambda$ conjugate to $\lambda$.  For $SU(3)$, the 
IR $(N,M)$ is the conjugate of $(M,N)$.  Define basis states
\begin{eqnarray}
\vert^\lambda_\alpha \rangle_c							\label{eq:7.9}
\end{eqnarray}
transforming like
\begin{eqnarray}
g : \vert^\lambda_\alpha \rangle_c \rightarrow \sum_\beta (D(g)^\lambda_
{\alpha \beta} )^* \vert^\lambda_\beta \rangle_c 			
\label{eq:7.10}
\end{eqnarray}
This means,
\begin{eqnarray}
g: \sum_\alpha \vert^\lambda_\alpha \rangle \vert^\lambda_\alpha \rangle_c
\rightarrow \sum_\alpha (\sum_\beta(D(g))^\lambda_{\alpha \beta}
\vert^\lambda_\beta \rangle ) (\sum_\gamma (D(g)^\lambda_{\alpha\gamma})^*
\vert^\lambda_\gamma  \rangle_c)\nonumber \\
= \sum_{\beta , \gamma} (D^{\lambda\dagger}(g)
D(g)^{\lambda} )_{\gamma \beta} \vert^\lambda_\beta \rangle
\vert^\lambda_\gamma \rangle_c = \sum_\alpha \vert^\lambda_\alpha
\rangle \vert^\lambda \alpha \rangle_c  
\label{eq:7.11}
 \end{eqnarray}
since $D(g)$ is unitary.
Therefore
\begin{eqnarray}
\frac{1}{\sqrt{d_\lambda}} \sum_\alpha \vert^\lambda_\alpha \rangle
\vert^\lambda_\alpha \rangle_c 							\label{eq:7.12}
\end{eqnarray}
(where $d_\lambda$ is the dimension of the IR $\lambda$) (i) is invariant
under the group; (ii) is the unique invariant vector in the direct product
space of $\lambda$ and $\bar \lambda$ (since there is one, and only one 
singlet in the decomposition of the direct product of $\lambda$ and 
$\bar\lambda$.); (iii) it has unit norm.
	Now $\vert^{\lambda"}_{\alpha"} \rangle^k$ in (eq:~\ref{eq:7.2}) 
transforms like the IR $\lambda"$ for every $k$.  Therefore,
\begin{eqnarray}
\frac{1}{\sqrt {d_\lambda}} \sum_\alpha \vert^{\lambda"}_{\alpha"} 
\rangle^k \vert^{\lambda"}_{\alpha"} \rangle_c 	\label{eq:7.13}
 \end{eqnarray}
is  an invariant with unit norm for every $k$.  Use this in
(eq:~\ref{eq:7.2}).  We get, in the direct product of three IRs, $\lambda, \lambda'$ 
and $ \lambda"$,
\begin{eqnarray}
\frac{1}{\sqrt{d_\lambda^"}} \sum_{\alpha,\alpha',\alpha"}
\langle^{\lambda \lambda'}_{\alpha \alpha'} \vert^{\lambda"}_{\alpha"}
\rangle^k \vert^\lambda_\alpha \rangle \vert^{\lambda'}_{\alpha'} \rangle
\vert^{\lambda"}_{\alpha"} \rangle_c						\label{eq:7.14}
\end{eqnarray}
(i) is, for each $k$, an invariant of unit norm,
(ii) are linearly independent vectors in the direct product space for 
various $k$'s. (This is because $\vert^{\lambda"}_{\alpha"} \rangle^k$ 
in (eq:~\ref{eq:7.2}) are linearly independent for various $k$.)
(iii) are the only invariant vectors in the direct product space of 
$\lambda, \lambda'$ and $  \lambda".$ (This is because (eq:~\ref{eq:7.12}) 
are the only invariant vectors formed from $\vert^{\lambda"}_{\alpha"} 
\rangle^k.$ We will rewrite (eq:~\ref{eq:7.14}) in the following way. 
Define the $3-G$ {\it symbols},
\begin{eqnarray}
[^{\lambda \lambda' \lambda"}_{\alpha \alpha' \alpha"}]_k =
\frac{1}{\sqrt{d_{\lambda^{"}}}} \langle^{\lambda \lambda'}_
{\alpha \alpha'} \vert^{\lambda"}_{\alpha"} \rangle^k_c	\label{eq:7.15}
\end{eqnarray}
Noting that
\begin{eqnarray}
(\vert^\lambda_\alpha \rangle_c)_c= \vert^\lambda_\alpha \rangle 
\label{eq:7.16}
\end{eqnarray}
We see that,
\begin{eqnarray} 
\sum_{\alpha,\alpha',\alpha"} [^{\lambda \lambda' \lambda"}_{\alpha 
\alpha' \alpha"} ]_k \vert^\lambda_\alpha \rangle \vert^{\lambda'}_
{\alpha'} \rangle \vert^{\lambda"}_{\alpha"} \rangle				\label{eq:7.17}
\end{eqnarray} 
give all linearly independent invariant vectors in the direct product 
space of three IRs $\lambda, \lambda'$ and $\lambda"$.  This gives the 
generalization of the definition (eq:~\ref{eq:3.2.28}) of $3j$ symbols.  
Thus the number of linearly independent invariant vectors and hence the 
$3-G$ symbols, is given by the outer multiplicity of $\lambda"$ in the 
direct product of $\lambda$ and $\lambda'$. For $SU(2)$ there is just one.

We now consider the normalization of the $3-G$ symbols.  The basis 
vectors of all three IR's in (eq:~\ref{eq:7.2}) are orthonormal.  
Therefore we get, 
\begin{eqnarray}
\sum_{\alpha\alpha'}\langle^{\lambda\lambda'}_{\alpha\alpha'}
\vert^{\lambda"}_{\alpha"}\rangle^k   \langle^{\lambda\lambda'}_
{\alpha\alpha'}\vert^{\lambda'''}_{\alpha'''}\rangle^{k'} = 
\delta_{kk^{'}}\delta_{\lambda''\lambda'''}\delta_{\alpha''\alpha'''}
\label{eq:7.18}
\end{eqnarray}
For $3-G$ symbols, this gives (see ~\ref{eq:7.15}),
\begin{eqnarray}
\sum_{\alpha\alpha'}\left [^{\lambda\lambda'\lambda''}_{\alpha\alpha'
\alpha''}\right ]_k   \left [^{\lambda\lambda'\lambda'''}_{\alpha\alpha'
\alpha'''}\right ]^*_{k'} = \frac{1}{d_{\lambda^{"}}}\delta_{kk^{'}}
\delta_{\lambda''\lambda'''}\delta_{\alpha''\alpha'''}
\label{eq:7.19}
\end{eqnarray}
\section{Invariants in the space ${\cal S}(M^1N^1)\otimes 
{\cal S}(M^2N^2) \otimes {\cal S}(M^3N^3)$}
The next step in obtaining a formula for the Clebsch-Gordan coefficients,
is to construct the invariants (eq:~\ref{eq:7.17}) using our realization 
of the basis vectors.  This corresponds to (eq:~\ref{eq:3.2.30}) in case 
of $SU(2)$.  However, the computation for $SU(3)$ is more complicated 
for the following reasons.

\begin{enumerate}
\item The variables $(z_1,z_2$,$\theta, w_1,w_2)$ that we use do not 
transform linearly under $SU(3)$ in contrast to the variables 
$(z_1, z_2)$ of the $SU(2)$ case (see Appendix A).

\item We can form more than one invariant in contrast to the $SU(2)$ case.
\end{enumerate}

	To analyze the situation we first ignore the constraint 
$\vec{w}\cdot\vec{z}=0$.  Consider three vector spaces, ${\cal P}
(M^1,N^1), {\cal P}(M^2N^2)$ and ${\cal P}(M^3N^3)$ built of polynomials 
in variables $(\vec{z}^1,\vec{w}^1), (\vec{z}^2,\vec{w}^2)$ and 
$(\vec{z}^3,\vec{w}^3)$ respectively.  ${\cal P}(M^a,N^a)$, $a=1,2 
{\mbox{ or }}3$ is the space of polynomials homogeneous of degree $M$ in 
$(z_1, z_2, z_3)$ and of degree $N$ in $(w_1, w_2, w_3)$, respectively.  
Invariant theory can be applied \cite{MM1,JP1,RM} to this situation. The 
result is that any invariant can be constructed out of the basic invariants,
\begin{eqnarray}
\begin{array}{ccc}
\vec{z}^1\cdot \vec{w}^1, & \vec{z}^1\cdot \vec{w}^2, & \vec{z}^1\cdot 
\vec{w}^3\cr \vec{z}^2\cdot \vec{w}^1, & \vec{z}^2\cdot \vec{w}^2, & 
\vec{z}^2\cdot \vec{w}^3\cr \vec{z}^3\cdot \vec{w}^1, & \vec{z}^3\cdot 
\vec{w}^2, & \vec{z}^3\cdot \vec{w}^3\label{eq:8.1} 
\end{array}
\end{eqnarray}
\begin{eqnarray}
\begin{array}{cc}
\vec{z}^1\cdot\vec{z}^2\times\vec{z}^3, & \vec{w}^1\cdot\vec{w}^2
\times\vec{w}^3
\end{array}
\label{eq:8.2}
\end{eqnarray}
Further, these invariants are not all independent, because 
\begin{eqnarray*}
(\vec{z}^1\cdot \vec{z}^2\times\vec z^3) (\vec{w}^1\cdot\vec{w}^2
\times\vec{w}^3)
\end{eqnarray*}
\begin{eqnarray*}
=\left |
\begin{array}{ccc}
z^1_1   & z^1_2   & z^1_3\\
z^2_1 & z^2_2 & z^2_3\\
z^3_1 & z^3_2 & z^3_3
\end{array}
\right |
\times \left |
\begin{array}{ccc}
w^1_1 & w^2_1 & w^3_1\\
w^1_2 & w^2_2 & w^3_2\\
w^1_3 & w^2_3 & w^3_3
\end{array}
\right |
\end{eqnarray*}

\begin{eqnarray}
=\left |
\begin{array}{ccc}
\vec{z}^1\cdot \vec{w}^1  & \vec{z}^1\cdot \vec{w}^2 & \vec{z}^1
\cdot \vec{w}^3\\ 
\vec{z}^2\cdot\vec{w}^1 & \vec{z}^2\cdot\vec{w}^2 & \vec{z}^2
\cdot\vec{w}^3\\ 
\vec{z}^3\cdot\vec{w}^1 & \vec{z}^3\cdot\vec{w}^2 & \vec{z}^3
\cdot\vec{w}^3 \\
\end{array}
\right |                              
\label{eq:8.3}
\end{eqnarray}
Thus we may remove either $ \vec{z}^1\cdot\vec{z}^2\times\vec{z}^3$ or 
$ \vec{w}^1\cdot\vec{w}^2\times\vec{w}^3$ 
from the list of basic invariants 
(eq:~\ref{eq:8.1}) and (eq:~\ref{eq:8.2}).  Any invariant is a function 
of the remaining  ten invariants.  To decide which of the two invariants 
(eq:~\ref{eq:8.3}) is to be kept, note that we are interested in 
polynomials in the three sets of six variables.  Therefore, we permit 
only positive integral powers of {\it{either}} 
$\vec{z}^1\cdot\vec{z}^2\times \vec{z}^3$ {\it{or}} 
$\vec{w}^1\cdot \vec{w}^2\times \vec{w}^3$ in addition 
to those of (eq:~\ref{eq:8.1}).  This gives all independent invariant 
polynomials.  We will not be repeating polynomials which are identical on 
using (eq:~\ref{eq:8.3}).  This is because if we were to replace 
$\vec{z}^1\cdot \vec{z}^2\times \vec{z}^2$ in a polynomial by 
$\vec{w}^1\cdot \vec{w}^2\times\vec{w}^3$ using (eq:~\ref{eq:8.3}), the 
latter would be appearing in the denominator.  Therefore it would
not coincide with any linear combination of the other polynomials we have
considered. 

 Thus a general invariant polynomial is a linear combination of the
following invariants.
\begin{eqnarray} 
{\cal P}(N(1,1),N(1,2),N(1,3),N(2,1),N(2,2),N(2,3),N(3,1),N(3,2),\nonumber \\
N(3,3),L) = (\vec{z}^1\cdot \vec{w}^1)^{N(1,1)} (\vec {z}^1\cdot \vec 
{w}^2)^{N(1,2)} (\vec {z}^1\cdot \vec {w}^3)^{N(1,3)} (\vec {z}^2\cdot 
\vec {w}^1)^{N(2,1)}\nonumber\\
(\vec {z}^2\cdot \vec {w}^2)^{N(2,2)} (\vec {z}^2\cdot \vec {w}^3)^
{N(2,3)} (\vec {z}^3\cdot \vec {w}^1)^{N(3,1)} (\vec {z}^3\cdot \vec 
{w}^2)^{N(3,2)} (\vec {z}^3\cdot \vec {w}^3)^{N(3,3)}\nonumber \\
\times \left ( {(\vec {z}^1\cdot \vec {z}^2\times \vec {z}^3)^L \quad 
{\mbox{or}} \quad (\vec {w}^1\cdot \vec {w}^2\times \vec {w}^3)^{-L}}
\right) \label{eq:8.4}
\end{eqnarray}
We have adopted the following notation. If $L$ is a positive(likewise  
negative) integer, the invariant has the term $(z^1\cdot z^2\times z^3)^L$
(likewise ($w^1\cdot w^2\times w^3)^{-L}))$. This way, both possibilities 
are labeled by a single integer $L$ taking both positive and negative 
values.  The other labels, $N(a, b), a,b = 1, 2$ {\mbox {or }}$ 3$ freely 
range over all non-negative integers. 

	The invariants in the space ${\cal P}(M^1,N^1) \otimes {\cal P}
(M^2,N^2)\otimes {\cal P}(M^3,N^3)$ have to be built from a linear 
combination of those invariants (eq:~\ref{eq:8.4}) for which
\begin{eqnarray}
\sum^3_{b=1}N(a,b)+L\epsilon(L)=M^a,\nonumber \\
\sum^3_{b=1}N(b,  a)+\vert L \vert\epsilon (-L)=N^a, {\mbox{  a  = 
}}1,2,3 \\
{\mbox{Here  }},\qquad  \epsilon (L)  = 1, \quad L \geq 0, \quad  \epsilon  =  -1, \quad L < 0
\label{eq:8.5}
\end{eqnarray}
For three given IRs, this gives 
six equations for ten unknowns $N(a,b) {\mbox{ and }} L$. Therefore, there 
are many independent invariants, in general.

	In order to obtain the $3-SU(3)$ symbols (Sec.7) we have to expand 
these invariants in terms of a realization of the $SU(3)$ basis vectors 
as polynomials in $\vec{z}^a$ and $\vec{w}^a$. The analogous procedure 
for $SU(2)$, discussed in section 2, is very simple because we have a 
simple monomial basis. All that is needed there is to collect coefficients 
in the Taylor's expansion. However in the $SU(3)$ case the basis in more 
complicated. In the realization using six complex variables used in earlier 
works \cite{MM1,JP1,JP2},\cite{RM}, the basis span only a subspace of the space 
of all polynomials. As a result, one is not even assured that a general 
invariant satisfying (eq:~\ref{eq:8.3}) can be expanded in the basis 
vectors. That it be so expandable imposes restriction on the coefficients 
of the linear combinations of the basic invariants(eq:~\ref{eq:8.4}). 
In fact each of the three IRs imposes one restriction on the linear 
combination. This effectively reduces the number of independent linear 
combinations that may be chosen. Any freedom that is left corresponds to 
the repeating IR's in the decomposition. We may hope that, requiring the 
invariant polynomial be an eigenstate of the 'chiral Casimir' 
 \cite{MM1,RM,GGH} uniquely determines invariant polynomial corresponding to each of 
the repeating IR's in the decomposition.

	In this way the $3-SU(3)$ symbols can be extracted - in principle. 
However, extracting an explicit formula for the symbols this way has not 
been possible.  The closest has been a formula with some undetermined 
coefficients \cite{RM}.

	We are able to overcome all these hurdles here for the following 
reasons. (i) Our basis is simpler (ii) The relevant invariants are completely 
determined (iii) We use our auxiliary measure to calculate the expansion 
coefficients (iv) We use generating function for the basis states and also 
for the invariants. This way calculations for all IRs are being done in 
one shot. Moreover, all calculations effectively reduce to Gaussian 
integrations.

	We now describe as to why the relevant invariants are completely 
determined in our basis.

	We are using the $5$-(complex)-dimensional subspace $\vec{w}\cdot
\vec{z} = 0$ in our basis. This constraint is invariant under $SU(3)$. 
Therefore, invariants in the larger six-dimensional space are also invariants 
when restricted to our subspace. However all invariants with non-zero 
$N(1,1), N(2,2)$ and $N(3,3)$ vanish identically because in our basis 
\begin{eqnarray}
\vec{w}^a\cdot\vec{z}^a=0, \quad a = 1, 2, 3 
\label{eq:8.6}
\end{eqnarray}
Thus the basic invariants are only,
\begin{eqnarray}
I(N(1,2), N(2,3), N(3,1), N(2,1), N(3,2), N(1,3), L) = 
(\vec z^1\cdot \vec w^2)^{N(1,2)}\nonumber\\
\times (\vec z^2\cdot \vec w^3)^{N(2,3)} (\vec z^3\cdot \vec w^1)^
{N(3,1)} (\vec z^2\cdot \vec w^1)^{N(2,1)}(\vec z^3\cdot \vec w^2)^
{N(3,2)} (\vec z^1\cdot \vec w^3)^{N(1,3)}\nonumber\\
\times \left [(\vec{z}^1\cdot \vec z^2\times \vec z^3)^L \quad {\mbox{or}} 
\quad (\vec w^1\cdot \vec w^2\times \vec w^3)^{-\vert L\vert}\right ] 
\label{eq:8.7}
\end{eqnarray}
We now simply have,
\begin{eqnarray} 
N(1,2)+ N(1,3) + L\epsilon(L) = M^1 \nonumber \\
N(2,3)+ N(2,1) + L\epsilon(L) = M^2 \nonumber \\
N(3,1)+ N(3,2)+ L\epsilon(L) = M^3 \nonumber \\
N(2,1)+ N(3,1)+ \vert L\vert\epsilon(-L) = N^1 \nonumber \\
N(3,2)+ N(1,2)+ \vert L\vert \epsilon(-L) = N^2 \nonumber \\
N(1,3)+ N(2,3)+ \vert L\vert\epsilon(-L) = N^3 
\label{eq:8.8}
\end{eqnarray}

Thus we are lead to the same equations for multiplicity as obtained in 
\cite{AGMS} from other considerations.  

Note that (eq:~\ref{eq:8.5}) implies 
\begin{eqnarray}
3L = \sum^3_{a=1}(M^a-N^a)
\label{eq:140a}
\end{eqnarray} 
Thus $L$ is completely determined by the three IR's chosen and is not an 
independent parameter.  It counts the number of invariants formed using 
det$({\cal U})=1$ condition.  It may be called the 'chirality number' of 
the invariant. 

	As a result, when we consider three given IRs $\lambda^1, 
\lambda^2$ and $\lambda^3$, we have just six non-negative integers 
$N(a,b), a\neq b $, constrained by five linearly independent equations 
(eq:~\ref{eq:8.8}).  Therefore, there can be more than one solution for 
the set $N(a,b)$.  This corresponds 
the multiplicity problem in the decomposition of the Kronecker product.  

It is possible, in principle, that there are more invariants in our
subspace which do not have an invariant extension into the larger space
 of $\vec{z}^a$ and $\vec{w}^a$ variables. We now argue that there are no 
further invariants constructed out of our basis vectors from three IRs.
In Sec.3, we obtained our basis vectors from basis vectors in the space 
of $\vec{z}$ and $\vec{w}$ variables by retaining those which are distinct 
in our subspace. Now a basis for {\it{all invariants}} in the larger space 
are cataloged by (eq:~\ref{eq:8.4}). Therefore, by simply imposing the 
constraints (eq:~\ref{eq:8.5}) and retaining non-trivial and distinct 
invariants, we get all invariants for our case.

	In earlier approaches one was not assured that an arbitrary linear
combination of the basic invariants (eq:~\ref{eq:8.4}) could be expanded 
in the basis vectors of the three IRs. We do not have such problem now. 
This is because our basis vectors form all polynomials in four variables 
$z_1,z_2,w_1$ and $w_2$ and all (positive or negative integral) powers of 
$z_3$. Thus, even the basic invariants (eq:~\ref{eq:8.4}) with constraints 
(eq:~\ref{eq:8.5})can be expanded in our basis. This means the following. 
We may simply regard each of basic invariants (eq:~\ref{eq:8.4}) as the 
(unnormalized) linearly independent invariant vectors in the direct product 
space of three IRs. For a given basic invariant (eq:~\ref{eq:8.4}), the 
three IRs of which it is composed is given by (eq:~\ref{eq:8.7}).

Thus the $3-SU(3)$ symbols are naturally labeled by the set of integers, 

\begin{enumerate}
\item $N(1,2)$, $N(2,3)$, $N(3,1)$, $N(2,1)$, $N(3,2)$, $N(1,3)$, $L$
\item $P^a$, $Q^a$, $R^a$, $S^a$, $U^a$, $V^a$, a=1,2,3.\\
In place of (2) we may use the quark-model labels,
\item $M^a$, $N^a$, $I^a$, $I^a_3$, $Y^a$, a=1,2,3.
\end{enumerate}

These labels are related by the constraints.  These constraints may be 
displayed as follows:

{\tiny{
\begin{eqnarray}
\left[\!\! 
\begin{array}{ccccccccccccccc}
&  N(1,2)  &&  N(1,3) && N(2,3) && N(2,1) && N(3,1) && N(3,2) \cr\\ 
N^2 &&  M^1  &&  N^3 && M^2 && N^1 && M^3 && N^2\cr\\ 
&& U^1,  R^1  && V^3 , S^3 && U^2,R^2  && V^1, S^1 && U^3, R^3 && 
V^2, S^2 \cr\\
&& P^1        && P^3       && P^2      && Q^1      && Q^3      && Q^2
\end{array}\!\!
\right]_L
\label{eq:140b}
\end{eqnarray}
}}
This is the analogue of $3-j$ symbol of $SU(2)$ prescribed as  an 
array  of nine integers , (eq:~\ref{eq:32B}).  As in that case, this  array 
is  highly  redundant.   This notation  is  nevertheless  useful, 
because the allowed values can be easily read off.
For convenience, we will adopt the following notation for the $3-SU(3)$ 
symbols:
\begin{eqnarray}
\left[ 
\begin{array}{ccccccc}
N(1,2) & N(2,3) & N(3,1) & L & N(1,3) & N(3,2) & N(2,1)\cr 
& M^1N^1 && M^2N^2 && M^3N^3\cr 
& I'I^{'}_3Y' && I^2I^{2}_3Y^2 && I^3I^3_3Y^3
\end{array}
\right]
\label{eq:140c}
\end{eqnarray}
\section{A generating function for the invariants}
Though our basis has simplified many aspects, it is still not simple enough 
to allow the expansion coefficients to be read off from the invariants 
(eq:~\ref{eq:8.8}).  In order to compute these coefficients we will use 
our auxiliary inner product.  We will also use generating functions (eq:
~\ref{eq:3.4.1}) for the basis states and also a generating function for 
the invariants.  This simplifies the computations drastically.  Moreover 
we are doing computations for all IRs in one shot.

	Define the generating function for the invariants.
\begin{eqnarray}
{\cal I}_\pm(j_{12}, j_{23}, j_{31}, j_{21}, j_{32}, j_{13}, j_\pm) \!=\! 
exp (j_{12} \vec{z}^1\cdot\vec{w}^2\!+\! j_{23} \vec{z}^2\cdot\vec{w}^3\!
+\! j_{31}\vec{z}^3\cdot\vec{w}^1\nonumber \\
+ j_{21}\vec{z}^2\cdot\vec{w}^1 + j_{32}\vec{z}^3\cdot\vec{w}^2 + j_{13}
\vec{z}^1\cdot\vec{w}^3 + (j_+ \vec{z}^1\cdot\vec{z}^2\times\vec{z}^3 
{\mbox{ or }} j_-\vec{w}^1\cdot\vec{w}^2\times\vec{w}^3))
\label{eq:9.1} 
\end{eqnarray}

	By a Taylor expansion in the sources $j$ we generate all
basic invariants (eq:~\ref{eq:8.8}).  Note that we use  $\it{ either }$ 
$j_+$ $\it{ or }$ $ j_-$ because we do not need both $z^1\cdot z^2\times 
z^3$ and $w^1\cdot w^2\times w^3$ together.

	The exponent in (eq:~\ref{eq:9.1}) is linear in each of the 
variables separately. This is an important feature which allows explicit 
computations.

	We may write,
\begin{eqnarray}
{\cal I}_\pm= \sum_{N(1,2)......., \vert L\vert}\quad j^{N(1,2)}_{12} 
j^{N(2,3)}_{23} j^{N(3,1)}_{31} j^{N(2,1)}_{21}
j^{N(3,2)}_{32} j^{N(1,3)}_{13} (j_\pm)^{\vert L\vert} \nonumber \\
\times \vert N(1,2), N(2,3), N(3,1), N(2,1), N(3,2), N(1,3),\pm 
\vert L \vert) \label{eq:9.2}
\end{eqnarray}
The ket's on the rhs of (eq:~\ref{eq:9.2}) are the unnormalized invariant 
vectors(eq:~\ref{eq:7.17}) in the direct product of three IRs.  For the 
corresponding normalized invariant vectors, we have 
\begin{eqnarray}
\vert\! N\!(\!1\!,\!2\!), N\!(\!2\!,\!3\!), N\!(\!3\!,\!1\!), 
N\!(\!2\!,\!1\!), N\!(\!3\!,\!2\!), N\!(\!1\!,\!3\!), \pm \vert L 
\vert > \equiv n^{{-1}/{2}}(N(1,2),\nonumber \\
N(2,3),N(3,1),N(2,1), N(3,2), N(1,3), \pm \vert L \vert ) \times \vert N
(1,2)........ ,\pm \vert L \vert )\nonumber \\
=\! \sum_{\{P^a,Q^a,R^a,S^a,U^a,V^a\}}\! 
{\tiny \left[ 
\begin{array}{ccccccc}
N(1,2) & N(2,3) & N(3,1) & L & N(1,3) & N(3,2) & N(2,1)\cr 
& M^1N^1 && M^2N^2 && M^3N^3\cr 
& I^1I^{1}_3Y^1 && I^2I^{2}_3Y^2 && I^3I^3_3Y^3
\end{array}
\right]}\nonumber \\
\times  \vert\! P^1\!,Q^1\!,R^1\!,S^1\!,U^1\!,V^1\!>\!
\vert P^2\!,Q^2\!,R^2\!,S^2\!,U^2\!,V^2\!>\!\vert\! P^3\!,Q^3\!,R^3\!,
S^3\!,U^3\!,V^3\!> \label{eq:9.3}
\end{eqnarray}
	where $n$ is the normalization factor. We have used our labeling 
(eq:~\ref{eq:3.4.3}) for the basis vectors.  The variables on the r.h.s. 
of (eq:~\ref{eq:9.3}) are related by (eq:~\ref{eq:3.15}), (eq:~\ref
{eq:3.16}) and (eq:~\ref{eq:3.18}) for each $a = 1,2,3$.

Consider the auxiliary inner product of ${\cal I}_\pm$ with 
${\mbox{\large\bf {g}}}^1$${\mbox{\large\bf {g}}}^2{\mbox{\large\bf{g}}}^3$
which is the product of the partition functions for basis vectors of the 
three IRs.  Using (eq:~\ref{eq:3.4.3}) and noting from (eq:~\ref{eq:6.2}) 
and (eq:~\ref{eq:6.9})
\begin{eqnarray}
(P,Q,R,S,U,V\vert\vert P',Q',R',S',U',V'>= N^{-\frac{1}{2}}(P,Q,R,S,U,V)
\nonumber \\
\times M (P,Q,R,S,U,V) \delta_{PP'}\delta_{QQ'}\delta_{RR'}\delta_{SS'}
\delta_{UU'}\delta_{VV'} \label{eq:9.4} 
\end{eqnarray}
We get,
\begin{eqnarray*} 
\int_\pm\equiv ({\mbox{\large\bf {g}}}^1{\mbox{\large\bf {g}}}^2{\mbox
{\large\bf {g}}}^3, {\cal I}_\pm)~~~~~~~~~~~~~~~~~~~~~~~~~~~~~
~~~~~~~~~~~~~~~~~~~~~~~~~~~~~~~~~~~~~~~
\end{eqnarray*}
\begin{eqnarray}
=\!\!\sum\!\!\prod^3_{a=1}\!\!\left ({\bar p}^{a^{P^{a}}}\!\!{\bar q}^
{a^{Q^{a}}}\!\!{\bar r}^{a^{R^{a}}}\!\!{\bar s}^{a^{S^{a}}}\!\!{\bar u}^
{a^{U^{a}}}\!\!{\bar v}^{a^{V^{a}}}  
\!N^{-\frac{1}{2}}\!(P^a\!\!,Q^a\!\!,R^a\!\!,S^a\!\!,U^a\!\!,V^a) 
M(P^a\!\!,Q^a\!\!,R^a\!\!,S^a\!\!,U^a\!\!,V^a)\right )\nonumber\\
 \times j^{N(1,2)}_{12} j^{N(2,3)}_{23} j^{N(3,1)}_{31}j^{N(2,1)}_{21}j^
{N(3,2)}_{32} j^{N(1,3)}_{13} (j_{\pm})^{\vert L \vert}~~~~~~~~~~~~~~
~~~~~~~~~~~~~~~~~~~~~~~~~~\nonumber \\
\times n^{+\frac{1}{2}}(N(1,2),N(2,3),N(3,1),N(2,1),N(3,2),N(1,3),\pm L)
~~~~~~~~~~~~~~~~~~~\nonumber \\
\times 
\left[
\begin{array}{ccccccc}
N(1,2) & N(2,3) & N(3,1) & L & N(1,3) & N(3,2) & N(2,1)\cr 
& M^1N^1 && M^2N^2 && M^3N^3\cr 
& I^1I^{1}_3Y^1 && I^2I^{2}_3Y^2 && I^3I^3_3Y^3
\end{array}
\right]~~~~~~
\label{eq:9.5}   
\end{eqnarray} 
	Thus the $3-SU(3)$ symbols can be computed by calculating
$({\mbox{\large\bf {g}}}_1{\mbox{\large\bf {g}}}_2{\mbox{\large\bf {g}}}_3,
{\cal {I}_\pm})$ with respect to the auxiliary inner product.
The normalization $n$ in (eq:~\ref{eq:9.5}) has to be evaluated separately, 
using (eq:~\ref{eq:7.19})
\section{Evaluation of $\int_+$}
We now evaluate $\int_+$.  We have,
\begin{eqnarray}
\int_{\pm} & = & \int d\mu^1
\int d\mu^2\int d\mu^3\quad {\mbox{\large\bf {g}}}^1\quad {\mbox{\large\
bf {g}}}^2\quad {\mbox{\large\bf {g}}}^3\quad {\cal I}_{\pm} 
\label{eq:10.1}
\end{eqnarray}
Here,
\begin{eqnarray}
\int\!\!d\mu^a\!\!=\!\!\int^{2\pi}_0\!\!\frac{d\theta^a}{2\pi}\!\!\int\!
\!\frac{d^2z^a_1}{\pi}\!\!\int\!\!\frac{d^2z^a_2}{\pi}\!\!\int\!\!\frac
{d^2w^a_1}{\pi}\!\!\int\!\!\frac{d^2w^a_2}{\pi}  exp(-\bar{z}^a_1z^a_1 -
\bar{z}^a_2z^a_2 - \bar{w}^a_1w^a_1 - \bar{w}^a_2w^a_2)~~~~
\label{eq:10.2}
\end{eqnarray}
\begin{eqnarray}
{\overline {\mbox{\large\bf {g}}}}^a = exp(\overline{r^a_pz^a_1 + 
r^a_qz^a_2 + s^a_qw^a_1 + s^a_pw^a_2 + u^ae^{i\theta^{a}} - 
v^ae^{-i\theta^{a}}(z^a_1w^a_1 + z^a_2w^a_2)}),\nonumber \\
{a=1,2,3}
\label{eq:10.3} 
\end{eqnarray}
\begin{eqnarray*}
\end{eqnarray*}

In (eq:~\ref{eq:10.3}), it is sufficient to use the 'mass shell' values 
(eq:~\ref{eq:3.4.9}) for the sources.  Further, in ${\cal I}_{\pm}$ 
( see (eq:~\ref{eq:9.1})), we have,
\begin{eqnarray}
\vec{z}^1\cdot \vec{w}^2 &=& z^1_1w^2_1 + z^1_2w^2_2 - exp({i\theta^{1} - 
i\theta^{2}})(z^2_1w^2_1 + z^2_2w^2_2),{\mbox{  etc.}}
\label{eq:10.4}
\end{eqnarray}
Also,
\begin{eqnarray}
\vec{z}^1\cdot \vec{z}^2 \times \vec{z}^3 &=& e^{i\theta^{1}}
(z^2_1z^3_2 - z^3_1z^2_2) + ({\mbox {cyclic}})
\label{eq:10.5}
\end{eqnarray}

Note that all exponents in $\int_+$ are bilinear in $z^a_{1,2}$ and 
$w^a_{1,2}$ variables.  Therefore these integrations can be explicitly 
done.  The only term that could have caused problems is $\vec{z}^1\cdot 
\vec{z}^2 \times z^3$ (eq:~\ref{eq:10.5}), which is related to the 
multiplicity problem and is apparently cubic.  However, since $z^a_3=
e^{i\theta^{a}} , a=1,2,3$ , this term is also bilinear.  On the other 
hand, $\vec{w}^1\cdot \vec{w}^2\times \vec{w}^3$ appearing in 
$\int_-$ is not bilinear after elimination of $w^a_3$.
We will handle this problem in sec.12.

	The form of $\int_+$ suggests the following operations.  First, 
dependence on $\theta^a, a=1,2,3$, can be completely transferred to
the sources.  Make a change of variables,
\begin{eqnarray}
z^a_1 \rightarrow e^{i\theta^{a}}z^a_1;\quad & \quad z^a_2 \rightarrow \
!e^{i\theta^{a}}\,\,\,z^a_2\nonumber \\
\bar z^a_1 \rightarrow e^{-i\theta^{a}}\bar z^a_1; \quad & \quad 
\bar z^a_2 \rightarrow \,e^{-i\theta^{a}}\bar z^a_2\nonumber \\
w^a_1 \rightarrow e^{-i\theta^{a}}w^a_1;\quad  & \quad w^a_2 \rightarrow \,
e^{-i\theta^{a}}w^a_2\nonumber \\ 
\bar w^a_1 \rightarrow e^{-i\theta^{a}}\bar w^a_1;\quad  & \quad \bar 
w^a_2 \rightarrow \,e^{-i\theta^{a}}\bar w^a_2
\label{eq:10.6}
\end{eqnarray}

The measure (eq:~\ref{eq:10.2}), remains unchanged.  On the other hand,
\begin{eqnarray}
{\mbox{\large\bf g}}^a \rightarrow exp (\overline{(r^a_pe^{i\theta^{a}})
z^a_1 + (r^a_qe^{i\theta^{a}})z^a_2 +
(s^a_qe^{-i\theta^{a}})w^a_1 + (s^a_pe^{-i\theta^{a}})w^a_2}\nonumber\\
 + \overline{ (u^ae^{i\theta^{a}}) - (v^ae^{-i\theta^{a}})
(z^a_1w^a_1 + z^a_2w^a_2)}
\label{eq:10.7}
\end{eqnarray}
Also, 
\begin{eqnarray}
j_{12}\vec{z}^1\cdot \vec{w}^2 \rightarrow j_{12}e^{i\theta^{1} 
- i\theta^{2}}(z^1_1w^2_1 + z^1_2w^2_2 -z^2_1w^2_1 -z^2_2w^2_2)						
\label{eq:10.8}
\end{eqnarray}
with similar changes for $j_{23}{\vec z}^2\cdot {\vec w}^3,\, j_{21}
{\vec z}^2\cdot {\vec w}^1$ etc.  Lastly, 
\begin{eqnarray}
j_+\vec{z}^1\cdot \vec{z}^2 \times \vec{z}^3 \rightarrow j_
+e^{i\theta^1 + i\theta^2 +
i\theta^3}\left ((z^2_1z^3_2-z^3_1z^2_2) + ({\mbox {cyclic}})\right )	
\label{eq:10.9}
\end{eqnarray}

	This change of variables makes integration over 
$\theta^a, a=1,2,3 $ 
very easy.  Indeed, these integrations only implement one additive
conservation law for each of the three IRs, since
\begin{eqnarray}
\int^{2\pi}_0 \frac{d\theta}{2\pi} e^{in\theta} = \delta_{n,0}
\label{eq:10.10}
\end{eqnarray}

The three kinds of charges for various sources as can be read off from 
(eq:~\ref{eq:10.7}) - (eq:~\ref{eq:10.9}) are displayed in Table 2.  In 
effect, each of $(z^a_1, z^a_2, z^a_3) $ is given $Q^a$ charge $+1$ and 
each of $(w^a_1,w^a_2,w^a_3)$ has $Q^a=-1.$  

The sources have the corresponding compensating charges.  As a consequence, 
it is not necessary to explicitly do $\theta^a $ integrations.  We may 
simply ignore the dependencies on $\theta^a$. After integration over 
$z^a$ and $w^a$ variables, we only keep polynomial in sources each term 
of which is neutral with respect to $Q^1, Q^2$ and $Q^3$ charges.\\
\begin{table}
\begin{centering}
\begin{tabular}{llllllll}
$Q^1=+1$ & $\bar s^1$ &$ \bar v^1$ &$ j_{12} $ &$ j_{13}$ &$ j_+$ \\
$Q^1 =-1$ & $ \bar r^1 $ & $ \bar u^1 $ & $ j_{21} $ & $ j_{31} $  \\
$Q^2 = +1$ & $ \bar s^2$ & $ \bar v^2$ & $ j_{23}$ & $ j_{21}$ & $ j_+$\\
$Q^2=-1 $ & $ \bar r^2 $ & $ \bar u^2 $ & $ j_{32} $ & $ j_{12}$ \\
$Q^3 = +1$ & $ \bar s^3$ & $ \bar v^3 $ & $ j_{31} $ & $ j_{32}$ & $ j_+$\\
$Q^3 =-1$ & $ \bar r^3$ & $\bar v^3$ & $j_{13}$ & $ j_{23}$   
\end{tabular}
\caption{$\bar p^a,\bar q^a$, a=1,2,3 do not carry any of these charges.}
\end{centering}
\end{table}
>From Table 2  and our definitions of quantum numbers, powers of the 
sources in the polynomials have to satisfy,
\begin{eqnarray}
\sum_{b\neq a}N(a,b) + L - \sum_{b \neq a}N(b,a) &=& 
R^a +U^a - S^a - V^a, ~~ a = 1,2,3. 
\label{eq:10.12}
\end{eqnarray}
>From (eq:~\ref{eq:3.15}) and (eq:~\ref{eq:8.8}), we notice that both 
sides of (eq:~\ref{eq:10.12}), is $M^a-N^a$.  Thus the $\theta$ integration 
is only implementing equality of $M^a-N^a$ as calculated using the 
invariants (eq:~\ref{eq:8.8}) and the states (eq:~\ref{eq:3.15}).  
However, we know that $M$ and $N$ evaluated in these two ways should each 
be separately equal.  This stronger equality should be a consequence of 
integration over the $z$ and $w$ variables.

	We now consider the integrations over the $z$ and $w$ variables.  
It is convenient to employ following matrix notation, 
\begin{eqnarray}
Z_1= \pmatrix { z^1_1 \cr z^2_1 \cr z^3_1}, \quad\bar Z_1= \pmatrix 
{ \bar z^1_1 \cr \bar z^2_1 \cr \bar z^3_1},\quad \bar R_p= \pmatrix 
{ \bar r^1_p \cr \bar r^2_p \cr \bar r^3_p}
\label{eq:10.13}
\end{eqnarray}
with similar notations for $Z_2$, $\bar{Z}_2$, $W_1$, $\bar{W}_1$, $W_2$, 
$\bar{W}_2$, $\bar{R}_q$, $\bar{S}_p$, $\bar{S}_q$.  Further, define,
\begin{eqnarray}
\bar{V} = \pmatrix{\bar{v}^1 & 0 & 0 \cr 0 & \bar{v}^2 & 0  
\cr 0 & 0 & \bar{v}^3 } 
\label{eq:10.14}
\end{eqnarray}
\begin{eqnarray}
J = \pmatrix{+j_{31}+j_{21} & -j_{12} & -j_{13} \cr -j_{21} & +j_{12}
+j_{32} & -j_{23} \cr -j_{31} & -j_{32} & +j_{23}+j_{13} }    
\label{eq:10.15}
\end{eqnarray}
\begin{eqnarray}
A = \pmatrix{0 & 1 & -1 \cr -1 & 0 & 1 \cr 1 & -1 & 0 }
\label{eq:10.16}
\end{eqnarray}

In terms of these objects the integral we need is, 
\begin{eqnarray}
\int\! \frac{d^2\!Z_1}{\pi^3}\! \frac{d^2\!Z_2}{\pi^3}\! \frac{d^2\!W_1}
{\pi^3}\! \frac{d^2\!W_2}{\pi^3}\!exp(\!-\!{\bar Z_1}^T\!Z_1\! -\! 
{\bar Z_2}^T\!Z_2\! -\! {\bar W_1}^T\!W_1\! -\! {\bar W_2}^T\!W_2\! 
+\! {\bar Z_1}^T\! \bar R_p\! + \!{\bar Z_2}^T \!\bar R_q\nonumber \\
+\! {\bar W_1}^T\! \bar S_q\! +\! {\bar W_2}^T\! \bar S_p\!-\! 
{\bar Z_1}^T\! \bar V\! {\bar W_1}\! -\! {\bar Z_2}^T\! \bar V \!
{\bar W_2}\! -\! {Z_1}^T\! J\! W_1\! -\! {Z_2}^T\! J\! W_2\! +\! j_+\!
{Z_1}^T\! A\! Z_2\! +\! \sum^3_{a=1}\!{\bar u}^a\!)
\label{eq:10.17}
\end{eqnarray}
where,
\begin{eqnarray}
\int\! \frac{d^2\!Z_1}{\pi^3}\! =\! \int\! \frac{d^2\!z^1_1}{\pi}\! 
\int\! \frac{d^2\!z^2_1}{\pi} \!\int\! \frac{d^2z^3_1}{\pi}        
\label{eq:10.18}
\end{eqnarray}
etc. We now use (eq:~\ref{eq:3.5.5}) in the following order.
Integrate over $W_1$ and $W_2$.  We get,
\begin{eqnarray}
\int \!\frac{d^2\!Z_1}{\pi^3}\! \frac{d^2\!Z_2}{\pi^3}\! exp\!(\!
-\!{\bar Z_1}^T\!Z_1\! -\! {\bar Z_2}^T\!Z_2\! +\! {\bar Z_1}^T\! 
\bar R_p\! +\! {\bar Z_2}^T\! \bar R_q +\! Z_1^T \!J\! \bar V \!
{\bar Z_1}\! +\! Z_2^T\! J\! \bar V\!\bar Z_2\!\nonumber \\
-\! Z_1^T\! J\! \bar S_q\! -\! Z_2^T\!J\bar S_p\! +\! j_+\!Z_1^T\!AZ_2\! 
+\! \sum^3_{a\!=\!1}\!{\bar u}^a\!)  
\label{eq:10.19}
\end{eqnarray}
Now integrating over $Z_2$, we get,
\begin{eqnarray}
\int \!\frac{d^2\!Z_1}{\pi^3}\!det\!(\!1\! -\! \bar V \!J^T\!)^{-\!1}\! 
exp\!(\!-\!\bar Z_1^T\!(\!1\! -\! \bar V\! J^T\!)\!Z_1\! +\! 
\bar Z_1^T\!\bar R_p\! -\! \bar S_q^T\! J^T\!Z_1\!\nonumber \\
-\! \bar S_p^T\!J^T\!(\!1\! -\! \bar V\!J^T\!)^{-\!1}\! \bar R_q\! +\! j_
+\!Z_1^T\!A\!(\!1\! -\! \bar V\!J^T\!)^{-\!1}\!\bar R_q + 
\sum^a_{a=1}{\bar u}^a)      \label{eq:10.20}
\end{eqnarray}
Final integration  over $Z_1$ gives,
\begin{eqnarray}
det(1 - \bar VJ^T)^{-2}exp(-\bar S_q^TJ^T(1 - \bar VJ^T)^{-1}\bar R_p 
-{\bar S}^T_pJ^T(1 - \bar V J^T)^{-1}\bar R_q\nonumber \\
+\,j_+{\bar R}^T_q(1-\bar V J^T)^{T^{-1}}A^T(1-\bar V J^T)^{-1}{\bar R}_p
+ \sum^a_{a=1}{\bar u}^a)   
\label{eq:10.21}
\end{eqnarray}
We rewrite this as, 
\begin{eqnarray}
exp(-{\bar R_p}^TB J\bar S_q - \bar R_q^TB J\bar S_p + j_+\bar R_p^TBAB^T
\bar R_q +2tr(ln\; B) +\sum^a_{a=1}{\bar u}^a)                   
\label{eq:10.22}
\end{eqnarray}
where,
\begin{eqnarray}
B = (1 - J\bar V)^{-1} = \sum^\infty_{n=0}\frac{(J\bar V)^n}{n!}
\label{eq:10.23}
\end{eqnarray}
\section{Algebraic formula for $3-SU(3)$ coefficients when $L\geq 0$ }
	To get $3-SU(3)$ symbols for a given set of three IRs, we have 
to expand (eq:~\ref{eq:10.22}) in powers of the various sources.
	On mass shell we have,
\begin{eqnarray}
- {\bar R_p}^TBJ\bar S_q \!-\! \bar R^T_qBJ\bar S_p
\!=\!(\bar r^1(BJ)_{12}{\bar s}^2 \!-\! \bar {r}^2(BJ)_{21}{\bar s}^1)
(\bar p^1{\bar q}^2-\bar q^1{\bar p}^2)+{\mbox {cyclic}}  
\label{eq:11.1}
\end{eqnarray}
	Notice that we always have combinations such as $(\bar p^1
{\bar q}^2 - \bar q^1{\bar p}^2)$ which are invariant under the (isospin) 
$SU(2)$ transformations of the sources.  This is to be expected because 
our measure is manifestly invariant under this subgroup.  Note also that 
the diagonal terms of the matrix $(BJ)$ do not appear on the r.h.s (this 
is because of the negative sign in $\bar s^a_q = -\bar s^a\bar q^a,  a=1,
2,3$ etc.)This is again required by $SU(2)$ invariance.  With such diagonal 
terms we would get terms like $\bar p^a \bar q^a$ which are not $SU(2)$ 
invariant.  We also have,
\begin{eqnarray}
{\bar R_p}^TBAB^T\bar R_q = \bar r^1{\bar r}^2({\bar p}^1{\bar q}^2 
- {\bar q}^1{\bar p}^2)(BAB^T)_{12} +{\mbox {cyclic}}  
\label{eq:11.2}
\end{eqnarray}
Again we get $SU(2)$ invariant combinations. The reason now is the 
antisymmetry of the matrix $A$ and hence of $BAB^T$.

	Explicitly inverting the $3\times 3$ matrix $(1-J\bar V)$, we 
obtain quite simple expressions for the relevant matrix elements:
\begin{eqnarray}
(BJ)_{12} = \Vert{B}\Vert (-j_{12} + j_{12}(j_{23}+j_{13})\bar {v}^3
+j_{13}\bar {v}^3j_{32})\nonumber \\
(BJ)_{21} = \Vert{B}\Vert (-j_{21} + j_{21}(j_{23}+j_{13})\bar {v}^3
+j_{23}\bar {v}^3j_{31})
\label{eq:11.3}
\end{eqnarray}
and corresponding cyclic expressions.  Here, 
\begin{eqnarray}
\Vert{B}^{-1}\Vert = {\mbox {det}}(1-J\bar V)~~~~~~~~~~~~~~~~~~~~~~~~~
~~~~~~~~~~~~~~~~~~~~~~~~~~~~~~~~~~~~~\nonumber \\
= 1-((j_{31}+j_{21})\bar v^1 + {\mbox {cyclic}}) +
((j_{31}j_{12}+j_{31}j_{32}+j_{21}j_{32})\bar v^1\bar {v}^2 
+{\mbox {cyclic}})
\label{eq:11.4}
\end{eqnarray}
Also notice that,
\begin{eqnarray}
A_{ij} = \epsilon_{ijk}c_k		
\label{eq:11.5}
\end{eqnarray}
where 
\begin{eqnarray}
\vec c= (c_k) = (1,1,1)
\label{eq:11.6}
\end{eqnarray}
Therefore,
\begin{eqnarray}
(BAB^T)_{il}= B_{ij}\epsilon_{jkm}c_mB_{lk} = \Vert{B}\Vert \epsilon_{iln}
c_m(B^{-1})_{mn}				
\label{eq:11.7}
\end{eqnarray}
Now,
\begin{eqnarray}
c_m(B^{-1})_{mn} = \sum^3_{m=1} (1 - J\bar V)_{mn} = 1 {\mbox 
{ for each }} n	\label{eq:11.8}
\end{eqnarray}
Thus,
\begin{eqnarray}
(BAB^T)_{12} = (BAB^T)_{23} = (BAB^T)_{31}=\Vert{B}\Vert         
\label{eq:11.9}
\end{eqnarray}

	Now we have an explicit expression for $\int_+$ :
\begin{eqnarray}
\int_+\! \sim\! \Vert\!{B}\!\Vert^2\! exp [ \Vert\!{B}\!\Vert\!\left 
((\bar{u}^1\!+\!j_+\!{\bar r}^1{\bar {r}}^2-{\bar r}^1j_{12}{\bar {s}}^2 \!+\! 
{\bar {r}}^2j_{21}{\bar s}^1+{\bar r}^1j_{12}{\bar {s}}^2(j_{23}+j_{13})
{\bar {v}}^3 \right. \nonumber \\
\left. - {\bar {r}}^2j_{21}{\bar s}^1(j_{23} +j_{13}){\bar {v}}^3
+ {\bar r}^1j_{13}{\bar {v}}^3j_{32}{\bar {s}}^2-{\bar {r}}^
2j_{23}{\bar {v}}^3j_{31}{\bar s}^1\right)\nonumber \\
\times({\bar   p}^1{\bar   {q}}^2-{\bar   {p}}^2{\bar   q}^1)   + 
{\mbox{(cyclic)}} ]			
\label{eq:11.10}
\end{eqnarray}
where $\sim$ means that we are supposed to keep only terms consistent 
with the conservation laws (eq:~\ref{eq:10.12}).

This form implies another conservation law.
Note that $j_{21}$ and $j_{31}$ always appear with the ${\bar s}^1$ or 
${\bar v}^1$.  Therefore,  
\begin{eqnarray}
N(2,1) + N(3,1) = S^1 + V^1							\label{eq:11.12}
\end{eqnarray}
which we expect because both sides equal $N^1$.  We have similar equations 
for $N^2$ and $N^3$ also.  Taken with (eq:~\ref{eq:10.12}) which are a 
consequence of $\theta^a, a=1,2,3$ integrations, we get separate 
conservations of $M^a$ and $N^a$ $a=1,2,3$ as computed from the states 
and from the invariants.

	We now change the r.h.s of (eq:~\ref{eq:11.10}) to a form which 
automatically gives the conservation law:
\begin{eqnarray}
N(1,2) + N(1,3) + L = R^1 + U^1					
\label{eq:11.13}
\end{eqnarray}
and corresponding corresponding cyclic expressions.  For this we remove
$exp(\!\sum^3_{a=1}\!{\bar u}^a)$ and insert ${\bar u}^1, {\bar{u}}^2, 
{\bar{u}}^3$ factors suitably
in the other terms of the exponent and in $\Vert{B}\Vert$ :
\begin{eqnarray}
\int_+ \!\rightarrow\! \Vert\!\tilde{B}\!\Vert^2\! exp\! 
[\Vert\tilde{B}\Vert ((j_+{\bar r}^1{\bar r}^2{\bar u}^3\! -
\!{\bar r}^1j_{12}{\bar s}^2\! +\! {\bar r}^2j_{21}{\bar s}^1
\!+\!{\bar r}^1j_{12}{\bar s}^2({\bar u}^2j_{23}{\bar v}^3\!
+\!{\bar u}^1j_{13}{\bar v}^3)\nonumber \\
\!-\!{\bar r}^2j_{21}{\bar s}^1({\bar u}^2j_{23}{\bar v}^3\!
+\!{\bar u}^1j_{13}{\bar v}^3)
\!+\! {\bar r}^1j_{13}{\bar v}^3{\bar u}^3j_{32}{\bar s}^2\! 
-\! {\bar r}^2j_{23}{\bar v}^3{\bar u}^3j_{31}{\bar s}^1)\nonumber \\
\times ({\bar p}^1{\bar q}^2\!-\!{\bar p}^2{\bar q}^1)\! 
+\! {\mbox{(cyclic)}}]
\label{eq:11.14}
\end{eqnarray}
where
\begin{eqnarray}
\Vert\!\tilde{B}\!\Vert\! =\! 1\!-\!({\bar u}^3j_{31}{\bar v}^1\!
+\!{\bar u}^2j_{21}{\bar v}^1\! +\! {\mbox{(cyclic))}}\! 
+\! ({\bar u}^3j_{31}{\bar v}^1{\bar u}^1j_{12}{\bar v}^2\!+\!{
\bar u}^3j_{31}{\bar v}^1{\bar u}^3j_{32}{\bar v}^2 \nonumber \\
+ {\bar u}^2j_{21}{\bar v}^1{\bar u}^3j_{32}{\bar v}^2 + {\mbox{(cyclic))}}
\label{eq:11.15}
\end{eqnarray}
Now $j_{12}$ and $j_{13}$ always appears with ${\bar r}^1$ or ${\bar u}^1$ 
except in the terms ${\bar r}^1{\bar r}^2{\bar u}^3 + {\mbox{(cyclic)}}$.  
The effect of these last terms is to provide monomials where the powers 
$R^a+U^a,   a=1,2,3$  are equal.  The net effect is to imply 
(eq:~\ref{eq:11.13}) and corresponding cyclic expressions where $L$ is given by the 
sum of the powers of ${\bar r}^1{\bar r}^2{\bar u}^3, {\bar r}^2
{\bar r}^3{\bar u}^1$ and ${\bar r}^3{\bar r}^1{\bar u}^2$.  Thus the 
r.h.s of (eq:~\ref{eq:11.14}) automatically takes care of conservations 
laws, (eq:~\ref{eq:11.12}) and (eq:~\ref{eq:11.13}) and corresponding 
cyclic expressions.
It also gives $\int_+$ exactly except for the additional factor
\begin{eqnarray}
\frac{1}{U^1! U^2! U^3!}		
\label{eq:11.16}
\end{eqnarray}
(coming from $exp({\bar u}^1+{\bar u}^2+{\bar u}^3)$) to be associated 
with $({\bar u}^a)^{U^{a}},  a=1,2,3.$

	We have to now expand r.h.s of (eq:~\ref{eq:11.14}) in powers of 
the  various monomials in the exponent and determinant. This   is 
exactly analogous to the $SU(2)$ case (eq:\ref{eq:32A}).   As  in 
that  case, we have to collect all the terms contributing to  the 
monomial.   For  this we have first to adopt a notation  for  the 
powers of the monomials.  This is presented in Table 3. 

We  have  deliberately  adopted  this  notation  for  the  powers 
because,  the arguments in the symbols uniquely characterize  the 
term being considered.  
Thus  for  example $l(123)$ is associated  with  $(j_+)  (\bar{p}^1 
\bar{r}^1)  (\bar{q}^2 \bar{r}^2) (\bar{u}^3)$.  We have  such  variables 
associated with every permutation of $(123)$ arising from the term 
$j_+    \bar{r}^1   \bar{r}^2\bar{u}^3   (\bar{p}^1\bar{q}^2    - 
\bar{p}^2\bar{q}^1) + $  cyclic
in  the  exponent  in (eq:~\ref{eq:11.14}).  In Table 4,  we  catalogue  all 
allowed arguments in our variables of Table 3.
The  advantage  of our notation is that we can  easily  trace  the 
terms that involve a given source $j_{12}, \bar{p}^1$ etc.  Thus for 
example, we can write the conservation laws in a compact form  as 
below.  By collecting the powers of each source, we get,
\begin{eqnarray}
P^\alpha = \sum(l(\alpha--) + k(\alpha---) + m(\alpha-----) 
+ n(\alpha-----));~~~\nonumber \\
Q^\alpha = \sum(l(-\alpha-) + k(-\alpha--) + m(-\alpha----) 
+ n(-\alpha----));~~~~\nonumber \\
R^\alpha = \sum(l(\alpha--) + l(-\alpha-) +k(--\alpha-) ~~~~~\nonumber\\
+ m(--\alpha---) + n(--\alpha---));~~~~~~~~~~~~~~~~~~~~~~~~~\nonumber \\
S^\alpha = \sum(k(---\alpha) +  m(---\alpha--) + n(---\alpha--));~~~~~~
~~~~~~~~~~~~~\nonumber \\
U^\alpha = \sum(l(--\alpha)+  m(----\alpha-) + n(----\alpha-)) + e(\alpha-)
~~~~~~~~~~~\nonumber \\
+ f(\alpha---) + f(--\alpha-) + 2g(\alpha\beta\gamma));~~~~~~~~~~~~~~~~~
~~~~~~~~~~~~~~ \nonumber \\
V^\alpha = \sum(\! m(-----\alpha) + n(-----\alpha)) + e(-\alpha)~~~~~~~
~~~~~~~~~~~~~~~~~~~\nonumber \\
+ f(-\alpha--) + f(---\alpha) + g(-\alpha-) +g(--\alpha);~~~\nonumber \\
L = \sum l(---);~~~~~~~~~~~~~~~~~~~~~~~~~~~~~~~~~~~~~~~~~~~~~\nonumber\\
N(\alpha,\beta)  = \sum( k(--\alpha\beta) + m(--\alpha\beta--)) + 
m(----\alpha\beta)~~~~~~~~~~~\nonumber \\
+ n(--\alpha\beta--) + n(---\alpha\beta)+e(\alpha\beta)+f(\alpha\beta--)
\nonumber~~~~\\
+f(--\alpha\beta)+g(\alpha\beta-)+g(\alpha-\beta));~~~~~~~~~~~~~~~~~~~~ 
\label{eq:11.17}
\end{eqnarray}
Here we have used the following notation : $\sum$ stands for summation 
over all allowed arguments in the blank spaces.

Note that (eq:~\ref{eq:11.16}) and (eq:~\ref{eq:11.17}) express the 
non-negative integers,\\ $P^\alpha$,...,$V^\alpha , N(\alpha,\beta)$ 
and $L, \alpha, \beta = 1,2$ or $3$\\
in terms of yet other non-negative integers of Table 3. It is easy to see 
that the constraint $P^\alpha + Q^\alpha = R^\alpha + S^\alpha$ for the 
labels of each IR $\alpha = 1,2$ or $3$ is satisfied.  Also the constraints 
(eq:~\ref{eq:11.12}) and (eq:~\ref{eq:11.13}) are satisfied as is seen 
from the positions of the labels and Table 3.

We may read off various conservation laws in the $3-SU(3)$ symbols from 
(eq:~\ref{eq:11.17}):
\begin{eqnarray}
\sum^3_{\alpha=1}P^\alpha = \sum^3_{\alpha=1}Q^\alpha 
\end{eqnarray}
 
This is valid because we are summing over all positions of $\alpha$ in 
the labels.  This is simply a statement of the conservation of $I_3$.

Similarly,
\begin{eqnarray}
\sum^3_{\alpha=1}U^\alpha = \sum^3_{\alpha=1}V^\alpha + L
\end{eqnarray}

This implies conservation of hypercharge as seen by rewriting as 
\begin{eqnarray}
\sum^3_{\alpha=1}(\frac{1}{3}(M^\alpha - N^\alpha)+V^\alpha - U^\alpha)=0
\end{eqnarray}
 
	When we expand the exponent in powers of each monomial, we 
collect a factor 
\begin{eqnarray}
\Vert \tilde B \Vert^{1+h}							\label{eq:11.18}
\end{eqnarray}
in (eq:~\ref{eq:11.14}).  Here
\begin{eqnarray}
-1+h= \sum(l() + k () + m() + n())		\label{eq:11.19}
 \end{eqnarray}
Alternately, note that the number of $\Vert \tilde{B} \Vert$ factors is 
the sum of the number of $s^\alpha$ variables $(\alpha=1,2,3)$ and the 
power of $j_+$, i.e,
\begin{eqnarray}
-1+h = L + S' + S^2 + S^3				\label{eq:11.20}
\end{eqnarray}
We may now apply the formula
\begin{eqnarray}
(1-\sum^K_{k=1}x_k)^{-1-h} = \sum^\infty_{n_1,n_2=0}
\frac{(h+\sum n_k)!} {h! \prod_k(n_k!)} \prod^K_{k=1} x^{n_k}_k
\label{eq:11.21}
\end{eqnarray}
to calculate coefficients of various monomials in (eq:~\ref{eq:11.18}) 
where $\Vert \tilde{B} \Vert$ is as in (eq:~\ref{eq:11.14}).  This gives 
the coefficient of the monomials in the expansion of the r.h.s. of 
(eq:~\ref{eq:11.14}) to be,
\begin{eqnarray}
\frac{1+\sum(l() + k() + m() + n() + e() + f() + g())!} {(1 \!+ \!\sum
(l() \!+\! k() \!+\! m() \!+\! n())!. \prod (l())! (k())! (m())! 
(n())!(e())!(f())!(g())!}\nonumber \\
\times (-1)^{\sum_s(k() + n()) + \sum_Am() + \sum(f()+g())}
\label{eq:11.22}
\end{eqnarray}

Here $\sum$ and $\prod$ are over all possible arguments of the variables
indicated.  $\sum_S$ (respectively $\sum_A$) correspond to summations 
over only those arguments $(\alpha \beta\gamma\delta ...)$ such that 
$\alpha\beta$ is same as (respectively transposes of ) $\gamma\delta$.

	In order to get the coefficient of the monomial in $\int_+$, we 
have to multiply (eq:~\ref{eq:11.22}) by the factor (eq:~\ref{eq:11.16}).  
We may now extract the $3-SU(3)$ symbol from eqn. (eq:~\ref{eq:9.5}), by 
supplying factors of $N^\frac{1}{2}$ and $M^{-1}$ as required.  We get 
the $3-SU(3)$ symbol for $L\geq 0$.
\begin{eqnarray}
\left[ 
\begin{array}{ccccccc}
N(1,2) & N(2,3) & N(3,1) & L & N(1,3) & N(3,2) & N(2,1)\cr 
& M^1N^1 && M^2N^2 && M^3N^3\cr 
& I^1I^{1}_3Y^1 && I^2I^{2}_3Y^2 && I^3I^3_3Y^3
\end{array}
\right]
\nonumber\\
=n^{-1/2}(\!N\!(\!1,\!2\!)\!, N\!(\!2\!,\!3\!), N\!(\!3\!,\!1\!), 
N\!(\!2\!,\!1\!)\!, N\!(\!3\!,\!2\!)\!, N(1,3), L)\nonumber \\
\times [ \prod^3_{\alpha =1} \frac{P^{\alpha}!Q^{\alpha}!R^{\alpha}!
S^{\alpha}! U^{\alpha}!V^{\alpha}!(U^\alpha +2I^\alpha +1)!
(2I^\alpha +1)}{(V^\alpha + 2I^\alpha +1)!}]^{\frac{1}{2}}\nonumber \\
\times \sum_{e,f,g,k,l,m,n}\frac{(1+\sum e() + f() + g() + k() + l() + m() 
+ n())!}{(1+\sum(k() + l() + m() +n())! \prod (l()! \ldots n())!}\nonumber \\
\times (-1)^{\sum_S(k() + n()) + \sum_A m() + \sum (f() + g())}
\label{eq:11.23}
\end{eqnarray}
This is the exact analogue of (eq:~\ref{eq:32A}) of $SU(2)$.  The Clebsch 
- Gordan coefficients are presented as a sum over non -  negative 
integers which satisfy conditions (eq:~\ref{eq:11.17}).
\newline\\
\begin{table}
\caption{}
$$
\begin{array}{|c|c|c|c|c|c}\hline
{\mbox{Monomial}} & j_+{\bar p}^1{\bar q}^2{\bar r}^1{\bar r}^2
{\bar u}^3 & {\bar p}^2{\bar q}^1{\bar r}^1{\bar r}^2{\bar u}^3 &
{\bar p}^1{\bar q}^2{\bar r}^1j_{12}{\bar s}^2 &  {\bar p}^2{\bar q}^1
{\bar r}^1j_{12}{\bar s}^2  \\ \hline 
{\mbox{Order used in label}} & {\bar p}^1{\bar q}^2{\bar u}^3 & {\bar p}^2
{\bar q}^1{\bar u}^3 & {\bar p}^1{\bar q}^2{\bar r}^1{\bar s}^2 & 
{\bar p}^2{\bar q}^1{\bar r}^1{\bar s}^2 \\ \hline
{\mbox{Power}} & l(123) & l(213) & k(1212) & k(2112)\\
\hline  
\end{array}
$$
\linebreak
\linebreak
$$
\begin{array}{|c|c|c|c|c|c}\hline
{\bar p}^1{\bar q}^2{\bar r}^2j_{21}{\bar s}^1 & {\bar p}^1{\bar q}^2
{\bar r}^1j_{12}{\bar s}^2{\bar u}^2j_{23}{\bar v}^3 & 
{\bar p}^1{\bar q}^2{\bar r}^1j_{12}{\bar s}^2{\bar u}^1j_{13}{\bar v}^3 & 
{\bar p}^1{\bar q}^2{\bar r}^1j_{13}{\bar v}^3{\bar u}^3j_{32}
{\bar s}^2\\ \hline  
{\bar p}^1{\bar q}^2{\bar r}^2{\bar s}^1 & {\bar p}^1{\bar q}^2{\bar r}^1
{\bar s}^2{\bar u}^2{\bar v}^3 & {\bar p}^1{\bar q}^2{\bar r}^1{\bar s}^2
{\bar u}^1{\bar v}^3 & {\bar p}^1{\bar q}^2{\bar r}^1{\bar s}^2
{\bar u}^3{\bar v}^3 \\ \hline 
{k(1221)} & m(121223) & m(121213) & n(12123)\\ \hline  
\end{array}
$$
\linebreak
\linebreak
$$
\begin{array}{|c|c|c|c|c|c}\hline
{\bar u}^3j_{31}{\bar v}^1 & {\bar u}^2j_{21}{\bar v}^1 & {\bar u}^3
j_{31}{\bar v}^1{\bar u}^1j_{12}{\bar v}^2 & 
{\bar u}^2j_{21}{\bar v}^1{\bar u}^1j_{32}{\bar v}^2& {\bar u}^3j_{31}
{\bar v}^1{\bar u}^3j_{32}{\bar v}^2\\ \hline 
{\bar u}^3{\bar v}^1 & {\bar u}^2{\bar v}^1 & {\bar u}^3{\bar v}^1
{\bar u}^1{\bar v}^2 &{\bar u}^3{\bar v}^2{\bar u}^2{\bar v}^1 & 
{\bar u}^3{\bar u}^3{\bar v}^2{\bar v}^1 \\ \hline
e(31) & e(21) & f(3112) & f(3221) & g(321)\\  
\hline
\end{array}
$$
\end{table}
\newline
\begin{table}
\caption{}
$\alpha , \beta , \gamma \cdots = 1, 2 {\mbox { or }} 3$
\newline
\linebreak
\begin{tabular}{ll}
$l(\alpha\beta\gamma)$ & $:(\alpha\beta\gamma) 
{\mbox { is a permutation of }} (123)$\\
$k(\alpha\beta\gamma\delta)$ & $ :(\alpha\beta )
{\mbox{ is same or transpose of }} (\gamma\delta)$\\
$m(\alpha\beta\gamma\delta\epsilon\phi)$ & $:(\gamma\delta\phi) 
{\mbox{ is a permutation of }} (123);$\\
&: $(\alpha\beta {\mbox{ is same or transpose of }} \gamma\delta);$  \\
& $:\phi {\mbox{ is either }} \gamma {\mbox{ or }} \delta.$  \\
$n(\alpha\beta\gamma\delta\epsilon)$ & $ : (\gamma\delta\epsilon) 
{\mbox{is a permutation of of (123)}}$;\\
& $:(\alpha\beta {\mbox{ is same or transpose of }}\gamma\delta).$ \\ 
$e(\alpha\beta):\alpha\neq \beta.$~~~~~~~~~~~~~~~~~~~~~~  \\
$f(\alpha\beta\gamma\delta)$ & $:\beta \!=\! \gamma {\mbox { and }} 
(\!\alpha\beta\delta\!) {\mbox{ is a permutation }}$\\
& :${\mbox{of }} (\!123\!);$\\
$g(\alpha\beta\gamma)$ & $ :(\alpha\beta\gamma) 
{\mbox{ only even permutation of }} (123)$\\
\end{tabular}
\end{table}
\section{Discussion}
In our calculations for $SU(3)$ we have ignored the questions involving 
the choice of phases till now.  We will now make a careful analysis.  
With our definition (eq:~\ref{eq:3.4.3}) of the unnormalized basis vectors, 
note that the $\underline{3}^*$ is represented by $(-w_1, w_2, w_3)$ upto 
a normalization constant .  Our computations of the 
relative normalization, (eq:~\ref{eq:6.30})-(eq:~\ref{eq:6.31}) fixes only 
the magnitude of the normalizations.   The phases may be chosen arbitrarily 
as discussed in Sec. 2 for $SU(2)$ case.  Under a change of phases, 

\begin{eqnarray}
\vert E > \rightarrow exp(i\theta_E)\vert E>,
\end{eqnarray}
the representation matrix changes as, 
\begin{eqnarray}
(D(g))_{EE'}\rightarrow e^{i\theta_E} (D(g))_{EE'} e^{-i\theta_{E'}}
\end{eqnarray}
and remains unitary.

However one may want to make a choice of phases to rid the formulae of 
phases and relative signs if possible.  For instance, we may choose the 
phase to have $\underline{3}^*$ be represented by 
$\frac{1}{\sqrt{2}}(w_1, w_2, w_3)$.

Fortunately our basis vectors (eq:~\ref{eq:3.4.1}-\ref{eq:3.4.3}) have 
real coefficients even though they are represented by polynomials in 
complex variables.  Similarly, our invariants (eq:~\ref{eq:8.4}) have 
real coefficients.  Thus we are assured that the $3-SU(3)$ symbols are 
real as is checked in (eq:~\ref{eq:11.23}).

We now address the ambiguity in the phase of the Clebsch-Gordan 
coefficients.  In the definition (eq:~\ref{eq:7.2}), various coupled bas
is vectors $\vert\lambda" \alpha">^k$, are required to transform as an 
IR($\lambda"$).  This fixes phases of all $3-SU(3)$ symbols except for 
an overall phase for each $\lambda"$ and $k$.

In previous sections we calculated $3-SU(3)$ symbols only for $L\geq 0$ 
case. With our choice of regarding $w_3$ as a dependent variable, the 
relevant integrations could be explicitly computed in this case. We now 
show the $3-SU(3)$ symbols can be obtained for $L< 0$ also.  

In constructing the basis vectors, we could have as well chosen to 
eliminate $z_3$ instead of $w_3$.  We could have done all computations 
with this basis.  In this case, the integrations for $3-SU(3)$ symbols 
can be done explicitly for $L<0$ instead of $L>0$.

This is related to the invariance of the $3-SU(3)$ symbols under 
conjugation of the IRs involved.  (It is possible that the invariance is 
only upto an additional phase factor as may happen if the phases of the 
basis states are 
not chosen carefully).  In our formalism this invariance may be seen as 
follows.  Consider an expansion of an invariant (eq:~\ref{eq:8.4}) with 
$L\geq 0$ as a linear combination of the basis vectors 
(eq:~\ref{eq:3.4.1}) -(eq:~\ref{eq:3.4.3}).  Now consider an interchange 
$\vec z \leftrightarrow \vec w$ in this expansion.  The effect on the 
invariant is to change it to another invariant with,

\begin{eqnarray}
N(\alpha , \beta) \rightarrow N(\beta , \alpha), L \rightarrow -L
\label{eq:198}
\end{eqnarray}

This means,
\begin{eqnarray}
M^a\leftrightarrow N^a\quad a=1,2,3.
\label{eq:199}
\end{eqnarray}

The effect of ${\vec z}\leftrightarrow {\vec w}$ on the generating 
functions (eq:~\ref{eq:3.4.1}) is equivalent to the following changes:
\begin{eqnarray}
p^a\rightarrow -q^a, \quad q^a\rightarrow p^a, \quad r^a \rightarrow s^a, 
\quad s^a\rightarrow -r^a, \quad u^a \rightarrow v^a, 
\quad v^a \rightarrow u^a
\label{eq:200}
\end{eqnarray}

Thus the effect on the unnormlized basis states (eq:~\ref{eq:3.4.3}) is, 
\begin{eqnarray}
\vert PQRSUV) \rightarrow (-1)^{Q+R}\vert QPSRVU)
\label{eq:201}
\end{eqnarray}

In the usual notation,
\begin{eqnarray}
M^a\leftrightarrow N^a, \quad I^a\rightarrow I^a, \quad I^a_3 \rightarrow 
-I^a_3, \quad Y^a \rightarrow -Y^a
\label{eq:202}
\end{eqnarray}
and in addition an additional phase factor $(-1)^{Q+R}$ is picked up.

In  eq.( ~\ref{eq:11.14}),  note that the isospin  dependence  is  always 
contained  in  the  $SU(2)$  invariant  combination  of  sources, 
$(\bar{p}^1\bar{q}^2 - \bar{p}^2\bar{q}^1)$ etc.  As a consequence, it is 
possible to extract isoscalar factors also \cite{WBG1}.  However, we will not 
pursue this here.  
Equation( \ref{eq:11.14}) can be used, in principle, to extract Regge 
symmetries \cite{RT} of 
$SU(3)$  Clebsch  - Gordan coefficients.  We don't  attempt  this 
here.
\section{Summary of results}
     For easy accessibility we summarize our results in a self -
contained way here.

(i)Labels for the basic vectors.

Normalized basis vectors are denoted by,
$\vert M, N; P, Q, R, S, U, V\rangle$.
All labels are non-negative integers.
All IRs are uniquely labeled by $(M, N)$.
For  a given IR $(M, N)$, labels $(P, Q, R, S, U, V)$ take all non  - 
negative integral values subject to the constraints: 
$$
R + U = M, \quad S + V = N, \quad P + Q = R + S.
$$
The  allowed  values can be read off easily: $R$ takes  all  values 
from  $0$ to $M$, and $S$ from $0$ to $N$.  For a given  $R$  and 
$S$, $Q$ takes all values from $0$ to $R + S$.  

(ii) Explicit realization of the basis states.

Consider the coefficient of the monomial 
$$
p^P q^Q r^R s^S u^U v^V
$$
in
$$
exp(r(pz_1 + qz_2) + s(pw_2 - qw_1) + uz_3 + vw_3).
$$
Divide it by the normalization,
$$
\left[\frac{(U+2I+1)!(V+2I+1)!}{P!Q!R!S!U!V!(2I+1)}\right]^{1/2}
$$
This  then  provides an explicit realization  of  the  normalized 
basis  state $\vert{P Q R S U V}\rangle$

(iii)Generating  function  for the invariants :

All  Clebsch  -  Gordan coefficients can be  extracted  from  the 
following generating function of the invariants :

\begin{eqnarray*}
{\cal I}_\pm(j_{12}, j_{23}, j_{31}, j_{21}, j_{32}, j_{13}, j_\pm) \!=\! 
exp (j_{12} \vec{z}^1\cdot\vec{w}^2\!+\! j_{23} \vec{z}^2\cdot\vec{w}^3\!
+\! j_{31}\vec{z}^3\cdot\vec{w}^1\nonumber \\
+ j_{21}\vec{z}^2\cdot\vec{w}^1 + j_{32}\vec{z}^3\cdot\vec{w}^2 + j_{13}
\vec{z}^1\cdot\vec{w}^3 + (j_+ \vec{z}^1\cdot\vec{z}^2\times\vec{z}^3 
{\mbox{ or }} j_-\vec{w}^1\cdot\vec{w}^2\times\vec{w}^3))
\end{eqnarray*}

(iv) Multiplicity labels for the Clebsch - Gordan series:

For given three IRs, $(M', N'), (M^", N^"), (M''', N'''),$
Construct all solutions of 
\begin{eqnarray*} 
N(1,2)+ N(1,3) + L\epsilon(L) = M^1 \nonumber \\
N(2,3)+ N(2,1) + L\epsilon(L) = M^2 \nonumber \\
N(3,1)+ N(3,2)+ L\epsilon(L) = M^3 \nonumber \\
N(2,1)+ N(3,1)+ \vert L\vert\epsilon(-L) = N^1 \nonumber \\
N(3,2)+ N(1,2)+ \vert L\vert \epsilon(-L) = N^2 \nonumber \\
N(1,3)+ N(2,3)+ \vert L\vert\epsilon(-L) = N^3 
\end{eqnarray*}

\begin{eqnarray*}
3L = \sum^3_{a=1}(M^a-N^a)
\end{eqnarray*} 

Where $N(a, b)$, $a\neq b$ are non - negative integers.
They  provide unambiguous labels for the Clebsch - Gordan  series 
as follows.

For  given two IRs $(M, N)$ and $(M', N')$, construct all  $(M^", 
N^")$  for  which $N(a, b), a\ne b$ have  non - negative  integer 
solutions.   Then the reversed pair $(N", M")$ gives all  IRs  in 
the  Clebsch - Gordan series.  Multiplicity of solutions for  one 
$(M", N")$ provides the multiplicity of repeating IRs.  Therefore 
${N(a,b)}$ unambiguously provide the multiplicity labels.

(v) $3 - SU(3)$ symbol

$3 - G$ symbols are related to the Clebsch - Gordan  coefficients 
as  in  (eq:~\ref{eq:7.15}), and have more explicit  symmetry  than  the 
latter.  The $3 - SU(3)$ symbol is represented by,

{\tiny{
\begin{eqnarray*}
\left[\!\! 
\begin{array}{ccccccccccccccc}
&  N(1,2)  &&  N(1,3) && N(2,3) && N(2,1) && N(3,1) && N(3,2) \cr\\ 
N^2 &&  M^1  &&  N^3 && M^2 && N^1 && M^3 && N^2\cr\\ 
&& U^1,  R^1  && V^3 , S^3 && U^2,R^2  && V^1, S^1 && U^3, R^3 && 
V^2, S^2 \cr\\
&& P^1        && P^3       && P^2      && Q^1      && Q^3      && Q^2
\end{array}\!\!
\right]_L
\end{eqnarray*}
}}

Here  the top row specifies the multiplicity labels:  the  second 
and  third rows specify the usual complete set of labels for  the 
basis states of the three IRs.

(vi)Generating  function  for the $3 - SU(3)$ symbol  for  $L  > 
0$.

Extract coefficient of the monomial 
$$
j^{N(1,2)}_{12}j^{N(2,1)}_{21}j^{N(1,3)}_{13}j^{N(3,1)}_{31} 
j^{N(2,3)}_{23}   j^{N(3,2)}_{32}  j^L_+\prod^{3}_{\alpha  =   1} 
\bar{p}^{P_\alpha}\bar{q}^{Q_\alpha}\bar{r}^{R_\alpha}\bar{s}^{S_\alpha} 
\bar{u}^{U_\alpha} \bar{v}^{V_\alpha}
$$
in
\begin{eqnarray*}
\int_+ \!\rightarrow\! \Vert\!\tilde{B}\!\Vert^2\! exp\! 
[\Vert\tilde{B}\Vert ((j_+{\bar r}^1{\bar r}^2{\bar u}^3\! -
\!{\bar r}^1j_{12}{\bar s}^2\! +\! {\bar r}^2j_{21}{\bar s}^1
\!+\!{\bar r}^1j_{12}{\bar s}^2({\bar u}^2j_{23}{\bar v}^3\!
+\!{\bar u}^1j_{13}{\bar v}^3)\nonumber \\
\!-\!{\bar r}^2j_{21}{\bar s}^1({\bar u}^2j_{23}{\bar v}^3\!
+\!{\bar u}^1j_{13}{\bar v}^3)
\!+\! {\bar r}^1j_{13}{\bar v}^3{\bar u}^3j_{32}{\bar s}^2\! 
-\! {\bar r}^2j_{23}{\bar v}^3{\bar u}^3j_{31}{\bar s}^1)\nonumber \\
\times ({\bar p}^1{\bar q}^2\!-\!{\bar p}^2{\bar q}^1)\! 
+\! {\mbox{(cyclic)}}]
\end{eqnarray*}
Multiply by the factor
$$
\prod^3_{\alpha=1}\left[\frac{P^\alpha! Q^\alpha! R^\alpha!  R^\alpha! 
S^\alpha!  U^\alpha!  V^\alpha!  (U^\alpha  +  2I^\alpha  +   1)! 
(2I^\alpha + 1)}{(V^\alpha + 2I^\alpha + 1)!}\right]^{1/2}
$$
This  gives the $3 - SU(3)$ symbol up to an overall  normalization 
depending only on IRs involved.

(vii)Formula for $3 - SU(3)$ symbol for $L>0$.

We  have obtained an explicit analogue of the  Bargmann's  formula 
for the $3 - j$ symbol of $SU(2)$, (eq:~\ref{eq:32A}).
This formula for $3 - SU(3)$ symbols (for $L>0$) is presented  in 
(eq:~\ref{eq:11.13}).   The  notation  used for the  summation  variables  is 
defined in Tables 3 and 4 as explained in detail in Sec.11.

(viii)Generating function and formula for $3 - SU(3)$ symbol  for 
$L<0$ case

These can be obtained from those for $L>0$ by making the  changes 
indicated in (eq:~\ref{eq:198}-\ref{eq:202}).

{\Large\bf{Appendix A: Group action on the variables}}\\

In Sec. 5, we set $\vert z_3\vert = 1$ in constructing the auxiliary 
measure.  Thus the IRs are realized in the space of polynomials in 
$z_1, z_2, w_1, w_2$ and $e^{i\theta}$.  We clarify here the manner in 
which the group acts on these variables.

The action of the group on $\vec z$ and $\vec w$ is given by,
\begin{eqnarray}
{\cal U}:z_i \rightarrow {\cal U}_{ij}z_i \equiv z^{'}_i \qquad w_i 
= {\cal U}^*_{ij}w_j\equiv w_i'
\label{eq:(A.1)}
\end{eqnarray}

We have imposed the constraint, $\vec w \cdot \vec z = 0 $ and regarded 
$(z_1, z_2, z_3, w_1, w_2)$ as independent variables.  The action of the 
group on these variables is, 
\begin{eqnarray}  
{\cal U}:(z_1, z_2, z_3, w_1, w_2) \rightarrow (z_1', z_2', z_3',w_1', w_2')
\label{eq:(A.2)}
\end{eqnarray}
where the primed variables are defined in (eq:~\ref{eq:(A.1)}).

Now let us take $\vert z_3 \vert = 1$ i.e $z_3 = e^{i\theta}$.  However, 
under the action (eq:~\ref{eq:(A.2)}), $z_3 \neq 0$ in general:
\begin{eqnarray}
z_3' = \vert z_3' \vert exp(i\theta') 
\label{eq:(A.3)}
\end{eqnarray}

Now define, 
\begin{eqnarray}
z_1" = \frac{z_1'}{\vert z_3'\vert}, \quad z_2" 
= \frac{z_2'}{\vert z_3' \vert},\quad  w_1" 
= \frac{w_1'}{\vert z_3'\vert}, \quad w_2" = \frac{w_2'}{\vert z_3' \vert} 
\label{eq:(A.4)}
\end{eqnarray}

Then action of the group is defined by,
\begin{eqnarray}
{\cal U}:(z_1, z_2, w_1, w_2, \theta) \rightarrow (z_1", z_2",w_1", w_2", 
\theta') 
\label{eq:(A.5)}
\end{eqnarray}

This action is non-linear as seen from (eq:~\ref{eq:(A.1)}) and 
(eq:~\ref{eq:(A.4)}).  Inspite of this it serves our purpose as a 
calculating tool. 

Also the action is ambiguous whenever $ z^\prime_3 = 0 $, because 
$ \theta^\prime $ is then undefined. However, this does not pose a problem 
for us, because we use a generic situation in our calculations. 
$ z^{'}_3 = 0 $ is a set of measure zero in our space of variables, 
$ (z_1, z_2, z_3, w_1, w_2).$

\newpage

{\begin{center}
{\Large\bf{Appendix B: Examples}}
\end{center}

In this appendix we illustrate various aspects of our calculus with 
specific examples.  

Basis for $\underline {3}$, $\underline {3^*}$ and $8$ are prescribed in 
Tables 5, 6 and 7.  Quark model notation for these states is prescribed 
in first column.  The allowed quantum numbers $PQRSUV$ are computed from 
constraints (eq:~\ref{eq:3.15}), (eq:~\ref{eq:3.16}).  $I, I_3, Y$ are 
computed using (eq:~\ref{eq:3.18}).  The unnormalized basis states 
$\vert PQRSUV)$ are computed from (eq:~\ref{eq:3.4.1}) and 
(eq:~\ref{eq:3.4.3}).  The Gelfand-Zetlin normalizations $N^{1/2}(E)$ 
are given by (eq:~\ref{eq:6.32}).  Note that they all agree with earlier 
calculations.
\begin{table}
\begin{center}
\caption{$\underline{3}(M=1, N=0)$}
\begin{tabular}{||c|c|c|c|c|c|c|c|c|c|c|c||} \hline\\
- & $P$ & $Q$ & $R$ & $S$ & $U$ & $V$ & $I$ & $I_3$ & $Y$ & 
$\vert PQRSUV)$ & $N^{1/2}$\\\hline 
$u$ & $1$ & $0$ &  $1$ & $0$ & $0$ & $0$ & $1/2$ & $1/2$ & $1/3$ & 
$z_1$ & $\sqrt{2}$ \\ \hline 
$d$ & $0$ & $1$ &  $1$ & $0$ & $0$ & $0$ & $1/2$ & $-1/2$ & $1/3$ & 
$z_2$ & $\sqrt{2}$ \\ \hline 
$s$ & $0$ & $0$ &  $0$ & $0$ & $1$ & $0$ & $0$ & $0$ & $-2/3$ & 
$z_3$ & $\sqrt{2}$ \\ \hline
\end{tabular}
\end{center}
\end{table}
\begin{table}
\begin{center}
\caption{$\underline{3^*}(M=0, N=1)$}
\begin{tabular}{||c|c|c|c|c|c|c|c|c|c|c|c||} \hline\\
- & $P$ & $Q$ & $R$ & $S$ & $U$ & $V$ & $I$ & $I_3$ & $Y$ & 
$\vert PQRSUV)$ & $N^{1/2}$\\\hline 
${\bar d}$ & $1$ & $0$ &  $0$ & $1$ & $0$ & $0$ & $1/2$ & $1/2$ & 
$-1/3$ & $w_2$ & $\sqrt{2}$ \\ \hline 
${\bar u}$ & $0$ & $1$ &  $0$ & $1$ & $0$ & $0$ & $1/2$ & $-1/2$ & 
$-1/3$ & $-w_1$ & $\sqrt{2}$ \\ \hline 
${\bar s}$ & $0$ & $0$ &  $0$ & $0$ & $0$ & $1$ & $0$ & $0$ & $1$ & 
$w_3$ & $\sqrt{2}$ \\ \hline
\end{tabular}\\
\vspace{0.5cm}
Note that $\vert I=1/2, I_3=-1/2, Y=-1/3>$ is represented by 
$-\frac{1}{2}w_1$.
\end{center}
\end{table}
\begin{table}
\begin{center}
\caption{$\underline{8}(M=1, N=1)$}
\small{
\begin{tabular}{||c|c|c|c|c|c|c|c|c|c|c|c||} \hline\\
- & $P$ & $Q$ & $R$ & $S$ & $U$ & $V$ & $I$ & $I_3$ & $Y$ & 
$\vert PQRSUV)$ & $N^{1/2}$\\\hline 
$\pi^+$ & $2$ & $0$ &  $1$ & $1$ & $0$ & $0$ & $1$ & $1$ & $0$ & 
$z_1w_2$ & $\sqrt{6}$ \\ \hline 
$\pi^0$ & $1$ & $1$ &  $1$ & $1$ & $0$ & $0$ & $1$ & $0$ & $0$ & 
$-z_1w_1+z_2w_2$ & $\sqrt{12}$ \\ \hline 
$\pi^-$ & $0$ & $2$ &  $1$ & $1$ & $0$ & $0$ & $1$ & $-1$ & $0$ & 
$-z_2w_1$ & $\sqrt{6}$ \\ \hline 
$K^+$ & $1$ & $0$ &  $1$ & $0$ & $0$ & $1$ & $1/2$ & $1/2$ & $1$ & 
$z_1w_3$ & $\sqrt{6}$ \\ \hline 
$K^0$ & $0$ & $1$ &  $1$ & $0$ & $0$ & $1$ & $1/2$ & $-1/2$ & $1$ & 
$z_2w_3$ & $\sqrt{6}$ \\ \hline 
$\bar K^0$ & $1$ & $0$ &  $0$ & $1$ & $1$ & $0$ & $1/2$ & $1/2$ & 
$-1$ & $w_2z_3$ & $\sqrt{6}$ \\ \hline 
$K^-$ & $0$ & $1$ &  $0$ & $1$ & $1$ & $0$ & $1/2$ & $-1/2$ & $-1$ & 
$-w_1z_3$ & $\sqrt{6}$ \\ \hline 
$\eta$ & $0$ & $0$ &  $0$ & $0$ & $1$ & $1$ & $0$ & $0$ & $0$ & 
$(\!{\small{z_3\!w_3\!=\!-\!z_1\!w_1\!-\!z_2\!w_2}}\!)$ & $2$ \\ \hline 
\end{tabular}
}
\end{center}
\end{table}
Note that we have generated only eight states even though we have not 
explicitly rewritten $w_3$ in terms of $z_1, w_1, z_2$ and $w_2$.  This 
is because of the specific way in which $p$ and $q$ enter in the 
generating function (eq:~\ref{eq:3.4.1}).

We now present explicit examples of our labeling (eq:~\ref{eq:8.8}) 
for multiplicity.       
\begin{table}
\begin{center}
\caption{}
\small{
\begin{tabular}{||c|c|c|c|c|c|c|c|c|c|c|c||} \hline\\
Invariant & $N\!(\!1\!,\!2\!)$ & $N\!(\!2\!,\!3\!)$ & $N\!(\!3\!,\!1\!)$ & 
$N\!(\!2\!,\!1\!)$ & $N\!(\!3\!,\!2\!)$ & $N\!(\!1\!,\!3\!)$ & $L$ & 
Invariant \\\hline 
$\underline{3}\times\underline{3}^*\times\underline{8}$ & $0$ & $0$ & $0$ & 
$0$ & $1$ & $1$ & $0$ & {\tiny $(z^1\cdot w^3)(z^3\cdot w^2)$} \\\hline 
$(\underline{8}\times\underline{8}\times\underline{8})_1$ & $1$ & $1$ & 
$1$ & $0$ & $0$ & $0$ & $0$ & {\tiny $(z^1\cdot w^2)(z^2\cdot w^3)
(z^3\cdot w^1)$} \\\hline 
$(\underline{8}\times\underline{8}\times\underline{8})_2$ & $0$ & $0$ & 
$0$ & $1$ & $1$ & $1$ & $0$ & {\tiny $(z^2\cdot w^1)(z^3\cdot w^2)
(z^1\cdot w^3)$} \\\hline 
$\underline{3}\times\underline{3}\times\underline{3}$ & $0$ & $0$ & $0$ 
& $0$ & $0$ & $0$ & $1$ & {\tiny $(z^1\cdot z^2\times z^3)$} \\\hline 
\end{tabular}
}
\end{center}
\end{table}
In simple cases we may extract the $3-SU(3)$ symbols easily from the 
invariants (Table 8) and the explicit representations (Tables 5, 6, 7) 
(upto an overall normalization $n$).  We present this for the case 
$\underline {3} \times  \underline {3}^* \times \underline {8}$.  The 
relevant invariant is 
$({\vec {z^1}}\cdot{\vec{w^3}})({\vec{z^3}}\cdot{\vec{w^2}}
\cdot{\vec{w^3}})({\vec{z^3}}\cdot{\vec{w^2}})$.  

Using tables 5, 6 and 7, we have,
\begin{eqnarray}
({\vec{z^1}}\cdot{\vec{w^3}})({\vec{z^3}}\cdot{\vec{w^2}})~~~~~~~~~~~~~~
~~~~~~~~~~~~~~~~~~~~~~~~~~~~~~~~~~~~~~~~~~~~~~~~~\nonumber \\
= \sqrt{12}\vert{u}>\vert{\bar u}>\vert\pi^0>+2\vert {u}>\vert{\bar u}>
\vert\eta> -2 \sqrt{6}\vert{u}>\vert{\bar d}>\vert\pi^->\nonumber\\
 - 2\sqrt{6} \vert u > \vert\bar s > \vert K^- > - 2\sqrt{6} \vert d > 
\vert\bar u > \vert\pi^+ > + \sqrt{12} \vert d > \vert\bar d > 
\vert\pi^0 >\nonumber\\
- 2 \vert d > \vert\bar d > \vert\eta > + 2\sqrt{6} \vert d > 
\vert\bar s > \vert\bar {K^0}> - 2\sqrt{6} \vert s > \vert\bar u > 
\vert K^+>\nonumber\\
+ 2\sqrt{6}\vert s > \vert\bar d> \vert K^0 >+4\vert s >\vert\bar s > 
\vert \eta >
\label{eq:I}
\end{eqnarray}

The coefficients are the $3-SU(3)$ symbols upto an overall normalization 
constant.  We may compare this with the calculation in the quark
model \cite{GM}.  
The relevant invariant is, $\bar Q M Q$, where
\begin{eqnarray}
{\bar Q}=\pmatrix{\bar u, & \bar d, & \bar s}~~~~~~~~~~~~~~~~~~~~~~
~~~~~~~~~~~~~~~~~~~~~~~~~~~\nonumber \\
M=\pmatrix{\frac{1}{\sqrt{2}}\pi^0+\frac{1}{\sqrt{6}}\eta  & \pi^+ & K^
+ \cr \pi^- & -\frac{1}{\sqrt{2}}\pi^0+\frac{1}{\sqrt{6}}\eta  & K^0 \cr
K^- & \bar {K^0} & -\sqrt{\frac{2}{3}}\eta},\qquad Q=\pmatrix{u \cr d\cr s}
\label{eq:II}
\end{eqnarray}
\begin{eqnarray}
{\bar Q}MQ = \frac{1}{\sqrt{2}}u\bar u \pi^0 + \frac{1}{\sqrt{6}} u
\bar u \eta + d\bar u \pi^+ + s\bar u K^+ + u\bar d\pi^- - \frac{1}
{\sqrt{2}}d \bar d\pi^0 \nonumber\\
+ \frac{1}{\sqrt{6}}d\bar d\eta + s\bar dK^0 + u\bar sK^- + d\bar s
\bar {K^0} - \sqrt{\frac{2}{3}} s\bar s \eta
\label{eq:III}
\end{eqnarray}
The coefficients match with those in (eq:~\ref{eq:I}), if we multiply by 
an overall normalization factor $2\sqrt{6}$ and make the following change 
in the phases of the states:
\begin{eqnarray*}
\pi^0 \rightarrow -\pi^0 ,\quad \pi^- \rightarrow -\pi^-
,\quad K^- \rightarrow -K^- ,\quad \bar u \rightarrow
-\bar u
\end{eqnarray*}

We may check that our formula for the $3-SU(3)$ coefficients reproduces 
these numbers.  Instead of enumerating all non-zero $m(------)$ etc for 
this situation, we will employ the formula (eq:~\ref{eq:11.14}-eq:~\ref
{eq:11.15}).  We have to collect only the terms
linear in $j_{13}$ and $j_{32}$ in, 
\begin{eqnarray}
\Vert\tilde B\Vert^2exp[\Vert \tilde B\Vert(-\bar r^3j_{32}\bar s^2 (\bar
p^3\bar q^2 - \bar p^2\bar q^3) - \bar r^{1}j_{13}\bar s^3 
(\bar p^{1}\bar q^3 -
\bar p^3\bar q^{1})\nonumber\\
-  \bar r^{1}j_{13}\bar s^3 \bar u^3j_{32}\bar v^2(\bar p^3\bar
q^{1}- \bar p^{1}\bar q^3) +  \bar r^{1}j_{13}\bar v^3 \bar u^3j_{32}
\bar s^2 (\bar p^{1}\bar q^2 - \bar p^2\bar q^{1}))
\label{eq:IV}
\end{eqnarray}
where the relevant piece of $\Vert\tilde B\Vert$ is, 
\begin{eqnarray*}
\Vert\tilde B\Vert = 1 + \bar u^3j_{32}\bar v^2 + \bar u^j_{13}\bar v^3 
+ \bar u^3j_{32}\bar v^2 \bar u^{1}j_{13}\bar v^3
\end{eqnarray*}
We get the coefficient of $j_{13}j_{32}$ to be, 
\begin{eqnarray}
(\bar r^{1}\bar p^{1})(\bar s^2\bar q^2)(\bar r^3\bar s^3\bar p^3\bar q^3) 
- (\bar r^{1}\bar q^{1})(\bar s^2\bar q^2)(\bar r^3\bar s^3(\bar p^3)^2) 
- (\bar r^{1}\bar p^{1})(\bar s^2\bar p^2)(\bar r^3\bar s^3(\bar q^3)^2)
\nonumber\\
+ (\bar r^{1}\bar q^{1})(\bar s^2\bar p^2)(\bar r^3\bar s^3\bar p^3
\bar q^3) + 2(\bar r^{1}\bar q^{1})(\bar v^2) (\bar s^3\bar u^3\bar p^3) 
- 2(\bar r^{1}\bar p^{1})(\bar v^2) (\bar s^3\bar u^3\bar q^3)\nonumber\\
+ (\bar r^{1}\bar p^{1})(\bar s^2\bar q^2)(\bar u^3\bar v^3) 
- (\bar r^{1}\bar q^{1})(\bar s^2\bar p^2)(\bar u^3\bar v^3) 
+ 2(\bar u^{1})(\bar v^2)(\bar u^3\bar v^3)\nonumber\\
- 3(\bar u^{1}) (\bar s^2\bar q^2) (\bar r^3\bar v^3\bar p^3) + 3(\bar
u^{1})(\bar s^2\bar p^2)(\bar r^3 \bar v^3\bar q^3)
\label{eq:V}
\end{eqnarray}

Supplying appropriate factors of $N^{\frac{1}{2}}M^{-1}$ for the states 
of the three IRs in each term, we reproduce the coefficients in 
(eq:~\ref{eq:I}).  For instance,
the last term corresponds to $s\bar dK^0$.  The factors of
$N^{\frac{1}{2}}M^{-1}$ for these states are, $\sqrt{2}$, $\sqrt{2}$ and
$\sqrt{\frac{2}{3}}$ respectively.  Therefore we get the coefficients 
to be $2\sqrt{6}$ which agrees with (eq:~\ref{eq:I}).


\begin{thebibliography}{99}
\bibitem{BAO} Barut A.O and R.Raczka, The Theory of Group 
Representations and Applications, Polish Scientific Publications, 1977.
\bibitem{ZDP}  Zelobenko D.P.,  Compact Lie Groups and their 
Representations, (American  Mathematical  Society,  Providence,  Rhode  
Island (1973)) pp123.
\bibitem{WBG1}  Wybourne B.G., Classical Groups for Physicists, 
Wiley-Interscience (1970).		
\bibitem{GR}  Gilmore R., Lie Groups, Lie Algebras and Some of Their 
Applications, Wiley, New York (1974).
\bibitem{HM}  Hamermesh M, Group Theory and its Applications to Physical 
Problems, Addison-Wesley (1962).
\bibitem{MFD2}  Murunaghan F.D., The Unitary and Rotation Groups, 
Spartan Books, Washington, D.C., 1962.		 
\bibitem{WEP}  Wigner E.P., Group Theory and its Applications to the 
Quantum Mechanics of Atomic Spectra, Academic Press (1959).
\bibitem{WH}  Weyl H., The Classical Groups.  Their Invariants and 
Representations, Princeton University Press (1946).		
\bibitem{RG1}  Racah G., Group Theory and Spectroscopy, Lect. Notes, 
pp. 102, Princeton. Preprint by CERN, 1961, no. 61-68 and 
J.I.N.R. no R-1864.
\bibitem{BH}  Boerner H., Representations of Classical Groups. 
(North-Holland Publishing Co., 1963).
\bibitem{LDE}  Littlewood, D.E., The Theory of Group Characters and 
Matrix Representations of Groups, Oxford University Press, 1950.
\bibitem{LDB}  Lischtenberg D.B., Unitary Symmetry and Elementary 
Particles, Academic Press, 1970.
\bibitem{SY}  Smorodinskii Ya.A and L.A. Shelepin, Usp. Fiz. Nauk 
(Sov.) 106 (1972).
\bibitem{BLC9} Biedenharn L.C. and J.D. Louck, in Angular Momentum in 
Quantum Physics: Theory and Applications (Encyclopedia in Mathematics 
and its Applications, Vol 8). 
\bibitem{MM1}  Moshinsky M., Rev. Mod. Phys. 34 (1962) 813 ; 
J. Math.  Phys.  4 (1963) 1128.
\bibitem{MAJ}  Macfarlane A.J., L.O. O'Raifeartaigh and P.S. Rao, 
J. Math. Phys. 8 (1967) 536.
\bibitem{GGH}  Gadiyar G.H. and H.S. Sharatchandra, J. Phys. A: Math. 
Gen. 25 (1992) L85.
\bibitem{BLC7} Biedenharn L.C., M.A. Lohe., and J.D.Louck., J. Math. 
Phys.26 (1985) 1455.
\bibitem{BLC8} Biedenharn L.C. and J.D. Louck, in: The Racah-Wigner 
Algebra in Quantum Theory, (Encyclopedia of Mathematics and its 
applications, ed. Gian-Carlo Rota, Addison-Wesley, 1981).
\bibitem{PB}  Preziori B., A.Simoni and B.Vitale, Nuovo Cimento, 
34 (1964) 1101.
\bibitem{JP1}  Jasselette P., Nucl. Phys. 1 (1967) 521, and 529. 
\bibitem{JP2}  Jasselette P., J. Phys. A: Math. Gen. 13 (1980) 2261.
\bibitem{RM}  Resnikoff M., J. Math. Phys. 8 (1967) 63. 
\bibitem{OMF} O'Reilly M.F., J. Math. Phys. 23 (1982) 2022.
\bibitem{AGMS} Anishetty Ramesh, G.H.Gadiyar, Manu Mathur and H.S.Sharatchandra, Phys. Lett. B271, 391 (1991).
\bibitem{GS}    Gadiyar G.H. and H.S. Sharatchandra, Algebraic Solution 
of the Littlewood -- Richardson Rule: Multiplicity in $SU(n)$ Clebsch-
Gordan Series. IMSc Preprint \# 93/18. 
\bibitem{BLC5} Biedenharn L.C. and D. Flath, Commun. Math. Phys. 
93 (1984) 143.
\bibitem{BLC6} Biedenharn L.C., R.A. Gustafson, M.A. Lohe, J.D. Louck, 
S.C. Milne, in Special Functions: Gr. Theor. Aspects and Applications, 
D.Reidel Publ.Co.,1984.
\bibitem{LJD1}  Louck J.D., American Journal of Physics. 38 (1970) 3.
\bibitem{LJD2}  Louck J.D., and L.C.Biedenharn, Adv.Q. Chem (1990?).
\bibitem{GIM5}  Gelfand I.M. and A.V. Zelevinskij, Funkts. Anal.
Prilozh. 18(3) (1984) 14, Reprinted in Collected Works, ibid.
\bibitem{BLC4} Biedenharn L.C., A.Giovannini, J.D. Louck., 
J.Math.Phys.8 (1967) 691.
\bibitem{BAJ1}  Bracken A.J, Commun. Math. Phys. 94 (1984) 371.
\bibitem{BAJ2}  Bracken A.J. and J.H. MacGibbon, J. Phys. A: Math. 
Gen. 17 (1984) 2581.
\bibitem{DJ}  Deenen J. and C. Quesne, J. Phys. A: Math. Gen. 
19 (1986) 3463.
\bibitem{DJP}  Draayer J.P. and Akiyama, J. Math. Phys. 14 (1973) 1904.
\bibitem{LB1}  Le Blanc R., and D.J. Rowe, J. Phys. A: Math. Gen. 19 
(1986) 2913.
\bibitem{LB2}  Le Blanc R., and D.J. Rowe, J. Phys. A: Math. Gen. 19 
(1986) 1111.
\bibitem{KWH} Klink W.H., and T.T. Tuong, J. Comp. Phys. 80 (1989) 453.
\bibitem{vdW}  van der Waerden. Group Theoretical Methods in Quantum 
Mechanics, Springer-Verlag (1932). 
\bibitem{CE}  Cartan E, The Theory of Spinors, Hermann, Paris, 1966.
\bibitem{KHA} Kramers H.A, Quantum Mechanics, North-Holland Publishing 
Co., 1958.
\bibitem{BHC}  Brinkman H.C., Applications of Spinor Invariants in 
Atomic Physics, North-Holland Publishing Co. (1956).
\bibitem{SJ}  Schwinger J., "On  Angular  Momentum",   US   Atomic   
Energy Commission NYO-3071, (1952) unpublished ; reprinted  in  Quantum  
Theory  of  Angular   Momentum,   ed. L.C.Biedenharn and H.Van Dam 
(Academic Press, 1969).
\bibitem{BIN} Bernstein I.N., I.M. Gelfand and S.I. Gelfand, Proc. 
Petrovskij Sem. 2 (1976) 3.  Reprinted in Izrail M. Gelfand. Collected 
Works Vol.II, (Springer Verlag 1988) pp 464.
\bibitem{KVP}  Karasev V.P., N.P. Shchelok, Quantum Mechanics and 
Statistical Methods Edited By: M.M. Sushchinskiy, Nova Science 
Publishers 1988, Proc. Lebdev Phys. Inst. - Academy of Sciences of 
the USSR.  Series Ed: N.G. Basov, Vol.173.
\bibitem{PJS2} Prakash J.S., H.S.Sharatchandra, J. Phys. A: 
Math. Gen. 26 (1993) 1625-1633.
\bibitem{LE}  Loebl E., Group Theory and its Applications, Vols.I, II, 
III.Academic Press.
\bibitem{KD}  Kleima D., Nucl. Phys. 70 (1965) 577.
\bibitem{GIM6}  Gelfand I.M, I. N.Bernstein, S.I. Gelfand, Models 
of Representations of Lie.  Groups, I.M. Gelfand Collected Papers 
Vol II, Springer-Verlag (1988), pp494.
\bibitem{PJS1}  Prakash  J.S. and H.S.Sharatchandra (1992),  
"Optimal  Boson  Calculus for $SU(3)$ : Unique Resolution of the 
Weight Lattice by Random Integers", imsc. preprint, No. imsc-92/05, 
to be published.
\bibitem{BL}  Banyai L, N.Marinsen, I.Raszillier and V.Rittenberg, 
Phys. Lett. 14 (1965) 156. 
\bibitem{BM}  Baranger M., and E. Vogt, Ed: Advances in Nuclear 
Physics Vol I, Plenum Press (1968).
\bibitem{BV}  Bargmann V. Rev. Mod. Phys. 34 (1962) 829.  
\bibitem{BMA} Beg  Mirza A and H.Ruegge, J. Math. Phys. 6 (1965) 677. 

\bibitem{BRE} Behrends R.E., J. Dreitlein, C. Fronsdal, and W. Lee, 
Rev. Mod. Phys. 34 (1962) 1.

\bibitem{BLC1} Biedenharn L.C., J. Math. Phys. 4 (1963) 436.
\bibitem{BLC2} Biedenharn L.C., Phys. Lett. 3 (1963) 2541.
\bibitem{BLC3} Biedenharn L.C., in Lects. in Theoretical Physics. Vol 5 
Ed. W.F 
Brittin, B.W.Downs and Joanne Downs - Inter Science Publishers (1963).
\bibitem{DPAM} Dirac P.A.M., Principles of Quantum Mechanics, Oxford 
University Press, 1928.  
\bibitem{EJP2}  Elliott J.P and M. Harvey, Proc. Roy. Society  
A272 (1963) 557.
\bibitem{RT}  Regge T., NUOVO CIM. 10 (1958) 544.  Reprinted  in  Quantum  
Theory of Angular Momentum, ed. L.C.Biedenharn and H.Van Dam 
(Academic Press, 1969).
\bibitem{GM}  Gell-Mann M., Y.Neeman, The Eightfold Way, Benjamin INC, 1964. 
\end{thebibliography}
\end{document}